\documentclass[a4paper,fleqn,usenatbib]{mnras}

\usepackage[T1]{fontenc}
\usepackage{ae,aecompl}

\usepackage{graphicx}	
\usepackage{amsmath}	
\usepackage{amssymb}	
\usepackage{longtable,multicol}

\usepackage{etoolbox}
\makeatletter
\patchcmd\@combinedblfloats{\box\@outputbox}{\unvbox\@outputbox}{}{%
   \errmessage{\noexpand\@combinedblfloats could not be patched}%
}%
 \makeatother

\title[ PS1 Lenses \& Pairs: Spectroscopy and shear in 2M1134$-$2103]{A Search for Gravitationally Lensed Quasars and Quasar Pairs in Pan-STARRS1: Spectroscopy and Sources of Shear in the Diamond 2M1134$-$2103}

\author[Rusu et al.]
{
Cristian E. Rusu,$^{1}$\thanks{Subaru Fellow; e-mail: \texttt{cerusu@naoj.org}}
Ciprian T. Berghea,$^{2}$ 
Christopher D. Fassnacht,$^{3}$   
 \newauthor
Anupreeta More,$^{4,5}$
Erica Seman,$^{2}$   
George J. Nelson$^{2}$ 
and Geoff C.-F. Chen$^{3}$
\\
$^1$Subaru Telescope, National Astronomical Observatory of Japan, 650 N Aohoku Pl, Hilo, HI 96720\\ 
$^2$U.S. Naval Observatory (USNO), 3450 Massachusetts Avenue NW, Washington, DC 20392, USA\\
$^3$Department of Physics, University of California, Davis, CA 95616, USA\\ 
$^{4}$Kavli IPMU (WPI), UTIAS, The University of Tokyo, Kashiwa, Chiba 277-8583, Japan\\
$^{5}$Inter-University Centre for Astronomy and Astrophysics, Post Bag 4,Ganeshkhind, Pune 410007, India\\
}

\date{Accepted XXX. Received YYY; in original form ZZZ}

\pubyear{2018}

\begin{document}
\label{firstpage}
\pagerange{\pageref{firstpage}--\pageref{lastpage}}
\maketitle

\begin{abstract}
We present results of a systematic search for gravitationally
lensed quasars in Pan-STARRS1. Our final sample of candidates comprises of 91 systems, not including 25 rediscovered lensed quasars and quasar pairs.
In the absence of spectroscopy to verify the
lensing nature of the candidates, the main sources of
contaminants are likely to be quasar
pairs, which we consider to be a byproduct of our work, and a smaller number of quasar$+$star associations. Amongst the independently
discovered quads is 2M1134$-$2103, for which we obtained spectroscopy
for the first time, finding a redshift of 2.77 for the quasar. There is
evidence for microlensing in at least one image. We perform detailed
mass modeling of this system using archival imaging data, and find that
the unusually large shear responsible for the diamond-like configuration
can be attributed mainly to a faint companion $\sim4\arcsec$ away, and
to a galaxy group/cluster $\sim30\arcsec$ away. We also set limits of
$z\sim0.5-1.5$ on the redshift of the lensing galaxy, based on its
brightness, the image separation of the lensed images, and an analysis of
the observed photometric flux ratios.
\end{abstract}

\begin{keywords}
gravitational lensing: strong -- quasars: individual: 2M1134$-$2103
\end{keywords}


\section{Introduction} 
\label{sec:intro} 

To date, $\sim60$ quadruple (quad) and $\sim200$ double gravitationally lensed quasars have been discovered\footnote{\citet{lemon19} have compiled an up-to-date list of known lensed quasars, maintained at \url{https://www.ast.cam.ac.uk/ioa/research/lensedquasars/}. Also, C. Lemon, private communication.}. Their value as probes of cosmology and astrophysics has been explored observationally for the past four decades \citep[e.g., see reviews by][]{claeskens02,treu16}, yet their number is still a limiting factor for many focused studies \citep[e.g.,][]{oguri12,schechter14,bonvin17}. We are currently in a post-Sloan Digital Sky Survey \citep[SDSS;][]{york00} era when the large ongoing imaging surveys such as the Panoramic Survey Telescope and Rapid Response System \citep[Pan-STARRS1, hereafter PS1;][]{chambers16}, the Dark Energy Survey \citep{flaugher15} and the Hyper Suprime-Cam Subaru Strategic Program \citep{aihara18} do not (yet) have a spectroscopic counterpart, making it difficult to identify lensed quasars. As a result, contemporary dedicated searches for lensed quasars rely on selecting their candidates by applying machine learning techniques such as artificial neural networks \citep[e.g.,][]{agnello15} or gaussian mixture models \citep[e.g.,][]{ostrovski17,williams17} to multi-filter photometric catalogues in conjunction with pixel-by-pixel pattern recognition; by looking for flux and position offsets between these surveys and \textit{Gaia} \citep[e.g.,][]{lemon17,agnellospiniello18}, including capitalizing on the superior $Gaia$ resolution to resolve blended sources \citep[e.g.,][]{lemon18,agnello18b,agnello18c,delchambre18} and combining multiple such methods \citep[e.g.,][]{spiniello18,lemon19}; by assessing the plausibility of valid lensing configurations on automatically detected sources \citep[e.g.,][]{chan15}; and/or by complementing these with variability information \citep[e.g.,][]{berghea17,kostrzewa18}. 

Encouraged by the serendipitous discovery by \citet{berghea17} of the first quad from PS1, PSOJ0147, we have begun a systematic search for lensed quasars in this survey, by cross-correlating sources with the parent AGN catalogue of \citet{secrest15}. As the first PS1 data was released in December 2016, mining it for lensed quasars has only recently begun \citep[e.g.,][]{ostrovski18}, making it likely that other lensed quasars, including bright, large separation quads, are yet to be found. Given the PS1 sky coverage and depth, \citet{oguri10b} estimate that PS1 contains $\sim2000$ lensed quasars, including 300 quads. 

Recently, \citet[][hereafter L18]{lucey18} have announced the discovery of a new bright, large-separation quad, 2M1134$-$2103. This was a serendipitous discovery, as part of a search for extended 2MASS \citep{skrutskie97} sources in the PS1 footprint, to include as targets for the Taipan Galaxy Survey \citep{dacunha17}. As part of our search, we have independently discovered this system. Here, we aim to present a more in-depth modeling of the archival imaging data, looking in particular to identify the cause for the unusually large shear inferred in L18. In addition, we present for the first time spectroscopic data for this system. 

The structure of this paper is as follows: in Section~\ref{sec:search}
we describe our search technique and a new sample of lensed quasars and quasar pair candidates.
In Section~\ref{sec:imag} we describe our analysis of the archival
imaging data of 2M1134$-$2103, and in Section~\ref{sec:spec} our newly
acquired spectroscopic data. In Section~\ref{sec:lens} we present our
mass modeling of 2M1134$-$2103, and provide plausible explanations for
the unusually large shear. We conclude in Section~\ref{sec:conc}. Where necessary, we use a flat cosmology with $\Omega_\Lambda=0.74$ and $h=0.72$. 

\section{A search for gravitationally lensed quasars in PS1}
\label{sec:search} 

\subsection{Selection based on catalogue cuts and visual inspection}
\label{sec:searchvis} 

PS1 is a wide-field imaging system with a 1.8 m telescope and 7.7 deg$^2$ field of view, located on the summit of Haleakala in the Hawaiian island of Maui. The 1.4 Gpixel camera consists of 60 CCDs with pixel size of 0.256 arcsec \citep{ona08, ton09}. The first PS1 data release includes both images and a photometry catalogue \citep{chambers16}. PS1 uses five SDSS-like filters (g$_{P1}$, r$_{P1}$, i$_{P1}$, z$_{P1}$, y$_{P1}$). The largest survey PS1 performs is the 3$\pi$ survey, covering the entire sky north of $-30\deg$ declination. 

As we did for PSOJ0147, we start our search with the AGN candidates catalog of \citet{secrest15}, based on two mid-infrared colors measured with the \textit{Wide-field Infrared Survey Explorer} \citep[$WISE$;][]{wright10}. We cross-correlate this catalog with the PS1 catalog\footnote{We use the version available on Vizier, \url{http://vizier.u-strasbg.fr/viz-bin/VizieR}, which contains fewer contaminants} \citep{flewelling16} using a $3\arcsec$ radius cone search and keep 79951 candidates which have at least two counterparts (step {\it i}). Next, we remove candidates within 15 degrees of the galactic plane, resulting in 64055 remaining sources ({\it ii}). We then impose a faint magnitude cut of $i=19.5$ on the closest counterpart, in order to eliminate spurious candidates. This results in 25493 sources remaining ({\it iii})\footnote{Following step {\it iii}, we explored using an additional step to eliminate globular clusters and similar crowded regions, by imposing the condition that there are no more than seven counterparts within $10\arcsec$ radius. This would have eliminated only 182 systems, all of which we have explored visually, making this step unnecessary.}. Finally, we impose that the two brightest sources in each system should be similar in color, removing the ones with $g-i$ differences larger than 1.5 mag and $i-y$ differences larger than 1.0 mag ({\it iv}). The final sample contains 18015 candidates. 

We chose these cuts in order to recover most of the known lenses at the intersection of PS1 and the \citet{secrest15} catalogue, while resulting in a number of candidates small enough to allow visual inspection. From an all-sky catalogue of $\sim260$ known lenses \citep{lemon19}, which we matched with the \citet{secrest15} catalogue to insure a match within $10\arcsec$, we found 45 lenses for which their \citet{secrest15} catalogue counterparts have at least 2 detections in PS1 within $3\arcsec$ (corresponding to step {\it i}). These are further reduced to 44 (step {\it ii}), 32 ({\it iii}), and 30 lenses ({\it iv})\footnote{In addition to these, two other lensed quasars survive our selection and grading process, but are not picked up by the cross-match with the catalogue of lenses because of differences in the reported coordinates: SDSS~J1320+1644 and SDSS~J1433+6007.}. In addition to the cross-match with the known catalogue of lensed quasars, we also looked for previously known non-lens systems, by cross-matching the coordinates of our candidates with the list of known sources from the SIMBAD Astronomical Database\footnote{\url{http://simbad.u-strasbg.fr/simbad/sim-fcoo}} and the NASA/IPAC Extragalactic Database\footnote{\url{http://ned.ipac.caltech.edu/?q=nearposn}}. 

We downloaded $30\arcsec\times30\arcsec$ postage stamp color JPEG images of the candidates using the PS1 cutout service\footnote{\url{http://hla.stsci.edu/fitscutcgi_interface.html}}, which were then inspected visually by three of the authors (CTB, ES and GJN). Pairs with separation $\lesssim$ a few arcsec between components (consistent with strong lensing by galaxies) and similar colors, triplets with a redder inner component, as well as quads with configurations consistent with canonical lensing configurations were kept. Finally, another three authors (CER, AM and GCFC) graded the remaining sample of 448 candidates. As is customary in the lens search community, they used the following grading system: 0: unlikely to be a lens; 1: possibly a lens candidate (satisfies only some criteria to be a lens); 2: probably a lens candidate (satisfies most criteria to be a lens); 3: almost certainly a lens (there is almost no doubt that this is a lens). We find 312 systems with an average grade $\geq1$, and discard the rest.

Out of the 312 candidates, we recover a total of 15 known lenses. Of these, 6 are quads: PS~J0147+4630 \citep{berghea17}, 2M~1134-2103 \citep{lucey18}, SDSS~J1433+6007 \citep{agnello18a}, GraL~J1537-3010 \citep{lemon19,delchambre18}, PS~J1606-2333 \citep{lemon18} and PS~J1721+8842 \citep{lemon18}, and 9 are doubles: DES~J0245-0556 \citep{agnello18b}, PS~J0259-2338 \citep{lemon18}, HE~1104-1805 \citep{wisotzki93}, J1206-2543 (Lemon et al. in prep.), SDSS~J1206+4332 \citep{oguri05}, SDSS~J1320+1644 \citep{rusu12}, ULAS~J1405+0959 \citep{jackson12}, SDSS~J1515+1511 \citep{inada14}, and J2212+3144 \citep{lemon19}. This means that at the grading stage we miss the cluster quad SDSS~J1004+4112 \citep{inada03}. In addition, at the initial visual inspection stage to produce the list for grading we miss the quad PG1115+080 \citep{weymann80} and 15 doubles: PS~J0028+0631 \citet{lemon18}, J0102+2445 \citep{lemon19}, Q0142-100 \citep{surdej87}, PS~J0949+4208 \citet{lemon18}, SDSS~J1001+5027 \citet{oguri05}, SDSS~J1313+5151 \citep{ofek07}, SDSS~J1349+1227 \citep{kayo10}, SDSS~J1442+4055 \citet{more16}, ULAS~J1527+0141 \citep{jackson12}, PS~J2124+1632 \citet{lemon18}, another double from Ostrovski et al. in prep. and four more doubles from Lemon et al. in prep. 

Our cross-match with known lenses shows that we are more efficient at recovering quads than doubles, which is to be expected, because typical quad configurations are easier to identify visually. We are also biased against large-separation lenses, due to our requirement to have at least two components within 3\arcsec. Since at the visual selection stage we miss 17/32 of the known lenses included in our cutouts, we expect the completeness of our sample of candidates, defined as the ratio of the number of gravitational lenses in the final sample to the true number of lenses in the cutouts, to be $\lesssim 50\%$. Most of these are missed at the initial visual inspection stage. This can be attributed to two factors: first, most of the missed systems are doubles with only two clearly visible components in the cutouts, and with noticeable color differences between the components. On the other hand, the authors who have inspected the 18015 candidates have no formal experience with gravitational lenses. When the authors with formal experience graded 11/17 missed lenses, 9 of these received an average grade $\geq1$.
 
  We note that other known quads with bright lensing galaxies, such as 2M1310-1714 (L18), are not included in our sample because the lens light contaminates the infrared colors that the \citet{secrest15} AGN catalog is based on. \citet{secrest15} note that the chance of misclassifying stars in the AGN catalogue is $\leq 0.041\%$, so we expect that the main contaminants to our list of candidates, after visual examination, will be quasar + star pairs as well as quasar pairs, as either physically associated binary quasars or projected chance alignments. Indeed, 93 of our candidates, the great majority of those with spectroscopic results in the literature, consist of at least one AGN. 
 
We note that we have typically given a grade of 1 to candidates consisting of object pairs without signs of additional emission, as long as the separation was not too large. This is for two reasons: first, the lensing galaxy may be too faint to detect, which is consistent with the large fraction of known doubles we miss. This fraction would undoubtedly be even higher if we chose to exclude these pairs. Second, because rather than focusing on producing the purest lensed quasar sample, we prefer to include in our sample binary quasars and quasar pairs, which are of interest to the AGN community, for example for studies of quasar triggering \citep[e.g.,][]{hopkins08}, and of the small-scale quasar-quasar correlation function \citep[e.g.,][]{hennawi06,kayo12}.
 
 \subsection{Removal of quasar-star pairs using $Gaia$}
\label{sec:searchgaia} 
 
The recent availability of the $Gaia$ mission \citep{gaia16}, and in particular of its second data release catalogue \citep[DR2;][]{gaia18a}, has resulted in wide application in the latest searches for lensed quasars, as demonstrated by the multitude of recent studies enumerated in Section~\ref{sec:intro}. Here, we capitalize on the astrometric quantities included in this catalogue in order to further prune our list of candidates. 

$Gaia$ DR2 includes $\sim1.7$ billion sources over the whole sky, with a limiting magnitude of $G\sim21$ \citep{gaia18a}. With a full width at half maximum (FWHM) of $\approx0.1\arcsec$ \citep{fabricius16}, $Gaia$ is effective at deblending close pairs and clusters of objects, down to $0.4\arcsec$ in DR2 \citep{arenou18}. Multi-epoch photometry has enabled the measurement of proper motions and parallaxes for $\sim360$ million sources, and the Astrometric Excess Noise \citep[AEN;][]{koposov17} provides a means of separating compact galaxies from point sources. Color information ($Rp-Bp$) is also available for $\sim1.4$ billion sources. 

We have cross-matched our candidates with the $Gaia$ DR2 catalogue, in order to identify the counterparts of both PS1 sources in each candidate (up to four sources, in case of quads). Of the 312 candidates, 307 have detections in $Gaia$, 291 of these have measured parallaxes and proper motions, and 283 have measured colors. Of their companions (i.e., the secondary component in the pair of each system, or the brightest secondary component in case of quads), the corresponding numbers are 291, 276 and 260. 

We use the proper motion as a classifier, in the form of the proper motion significance defined by \citet{lemon19}, $\sqrt{(pm_\mathrm{ra}/\sigma_{pm_\mathrm{ra}})^2+pm_\mathrm{dec}/\sigma_{pm_\mathrm{dec}})^2}$, which includes both celestial coordinates, and where $\sigma$ stands for the measured uncertainty. We adopt a limiting upper value of 5, which recovers $\sim95\%$ of known lensed quasar images (see Figure 1 in \citet{lemon19}). For the parallax $\varpi$, we use $\varpi/\sigma_\varpi\leq4$, corresponding to a $4\sigma$ limit, since the distribution of measured parallaxes is well approximated by a Gaussian \citep{gaia18b}. Finally, we use AEN $\leq4$, corresponding to the limit which separates best between lensed quasar images and galaxies, and recovers $\sim90\%$ of the former (see Figure 2 in \citet{lemon19}).\footnote{Our chosen limits recover almost all of our confirmed candidates: amongst our 25 spectroscopically confirmed lenses or quasar pairs, only 2, both quads with 4 detected components in $Gaia$, would be (partially) ruled out based on our $Gaia$ classifier: PSJ0147+4630 has one component with large parallax, and PSJ1606-2333 has one with large proper motion.}

\begin{figure}
	\includegraphics[width=\columnwidth]{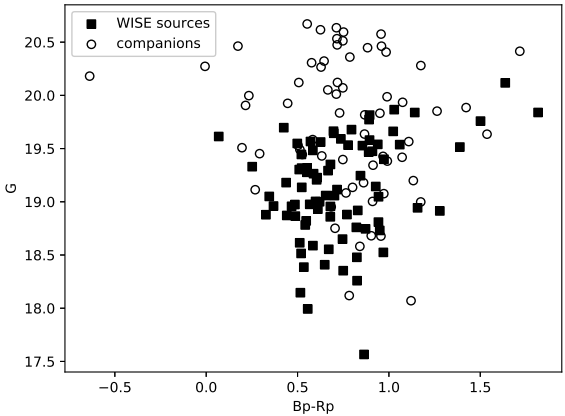}
	\caption{
$Gaia$ Color-magnitude plot of the main sources and companions with available colors, from the 91 surviving candidates.
}
    \label{fig:colormag}
\end{figure}

 Our final classification of the 312 candidates is: 91 surviving candidates yet unconfirmed (1 grade A, 4 grade B and 86 grade C), 25 confirmed systems (6 quads, 9 doubles\footnote{Note that it is presently unknown whether SDSS~J1320+1644, counted here as a double, is in fact a double or a quasar pair \citep{rusu12}.}, 10 quasar pairs), and 196 rejected candidates. We present our final sample of 91 gravitationally lensed quasar and quasar pair candidates in Table~\ref{tab:cand}, together with our comments based on visual inspection and $Gaia$ measurements. In Appendix~\ref{sec:appendix} we also list the already confirmed candidates, as well as the rejected ones. From this table it can be seen that the proper motion was the dominant classifier for the overwhelming majority of candidates. Finally, Figure~\ref{fig:colormag} shows that the companions have a similar distribution of $Gaia$ colors with the primary sources, but are typically fainter.
 
 \subsection{Expected sample purity}
\label{sec:searchpurity} 

In addition to the expected completeness, we also wish to estimate the purity of our sample, defined as the ratio of (number of gravitational lenses + quasar pairs)/(total number of sources in the sample). First, we perform a simple exercise where we estimate this number focusing only on the quasar pairs, and comparing the density of sources in a catalogue of point sources, and one of AGN. The idea is to estimate how many of the candidate source companions are expected to be AGN, as opposed to stars. We present the details of the computation in Appendix~\ref{sec:appendix1}. We arrive at a result of $\sim 4\%$.
  
The expected purity can be computed more directly using the subsample of candidates for which spectroscopic data is available in the literature. Out of 33 candidates which survive the $Gaia$-based cut and which can be either confirmed or ruled out based on the literature, 9 are doubles, 10 are quasar pairs, 6 are quads and 8 are galaxy+other, star+other or star+QSO (here ``other'' stands for non-QSO). For systems with $G\leq20$ for all components (the limit at which $Gaia$ is still relatively complete), these numbers are 7, 9, 5 and 5, respectively. This means that for $G\leq20$, if we ignore the quads (there is only one quad candidate in our final sample, and these systems are much easier to identify visually, leading to different selection), the purity for quasar pairs is $9/21\approx43\%$, and for doubles + quasar pairs it is $16/21\approx76\%$. Of course, care must be taken in interpreting this result, as the spectroscopic selection of these sources compiled from the literature is unknown.

How can the discrepant results of the two methods be reconciled? This is likely due to the known clustering of quasars, which leads to a significant enhancement of small-separation quasar-quasar pairs over expectations from uniform spatial distribution assumptions and catalogue density comparisons, and it means that the number we computed with that method must be interpreted as a lower limit. The quasar-quasar correlation function is predicted to produce an enhanced by a factor of $\sim100$ on small angular scales corresponding to quasar pairs \citep[e.g.,][and references therein]{peng99}. For our 91 candidates and confirmed quasar pairs we measure a median separation of $2.4\arcsec$, with a standard deviation of $0.53\arcsec$ (after removing 4 systems with separation $>4.5\arcsec$). \citet{kayo12} do indeed estimate an increase by a factor of $\sim200$ in the number of quasar pairs with separation typical for our candidates (physical scale $\sim20$ kpc), over the random expectation, based on a sample of binary quasars obtained as a byproduct of a search for gravitationally lensed quasars \citep{oguri06,inada12}. This is more than enough to explain the discrepancy. In fact, multiplying this number with the fraction of AGN to point sources found in our simple exercise suggests a purity of $\sim90\%$. This may be an overestimate, as there is a known discrepancy between the large number of predicted binary quasars \citep[e.g.,][]{hopkins06} and the smaller number of discovered ones \citep[e.g.,][]{hennawi10}. We adopt as our best estimate of the purity the $\sim76\%$ value measured above for quasar pairs + doubles, although we caution that this estimate might be biased due to the unknown spectroscopic selection, and applies only to $G\leq20$. If we remove the magnitude cut, based on the spectroscopic sample, this becomes $\sim70\%$.
 
 In the following sections, we focus on modeling the imaging and spectroscopic data of 2M1134$-$2103.

\begin{table*}
  \scriptsize
 \centering
 \begin{minipage}{\linewidth}
   \caption{Sample of gravitationally lensed quasar candidates and quasar pair candidates identified systematically from PS1}
  \begin{tabular}{@{}lrrclccl@{}}
  \hline 
Name [PS1~J...] & $\alpha$ & $\delta$ & \#Comp & $i$ & Sep. [\arcsec] & Rank & $G$ mags; notes \\ 
 \hline
000815-043634 & 2.061059  & $-4.609377$     & 2  &  18.57   & 2.4               & C & 18.82, 20.09; similar color p-l; both componentss negligible AEN, p, pm\\
003309-120520 & 8.287252  & $-12.088925$    & 3  &  18.10   & 6.8              & C & 18.93, 20.38; p-l (both negligible AEN, pm and p) + red inner component\\
004106+032726 & 10.273022 & $3.457205$      & 2  &  18.43   & 2.4              & C & 18.91, 20.36; p-l; includes SDSS z=1.282 QSO; \\
& &  &  &  & & & both negligible AEN, pm and p \\
004518+405433 & 11.325876 & $40.909217$     & 2  &  18.69   & 3.1               & C & 19.32, 18.85; similar color p-l; includes z=1.228 QSO \citep{huo13}; \\
& &  &  &  & & & both negligible AEN, pm and p \\
012221+291431 & 20.587958 & $29.242069$     & 2  &  18.30   & 2.4              & C & 18.41, 20.72; similar color p-l; one component negligible AEN, pm and p;\\
& &  &  &  & & & companion has no Gaia pm and p \\
012256+783855 & 20.733302 & $78.648546$     & 2  &  18.43   & 2.0               & C & 18.94, 18.99; similar color p-l; both negligible AEN, pm and p\\
012648+411136 & 21.698143 & $41.193204$     & 2  &  19.13   & 3.1              & C & 19.26, 20.15; similar color p-l; both negligible AEN, pm and p\\
013021+072516 & 22.585897 & $7.421231$      & 2  &  18.80   & 2.0               & C & 18.98, 19.65; p-l; one component negligible AEN, pm and p;\\
& &  &  &  & & & companion has no Gaia pm and p \\
014114-062740 & 25.307825 & $-6.461006$     & 2  &  19.05   & 2.4              & C & 20.79; similar color p-l; only one has Gaia data; negligible AEN, pm and p\\
014455+271137 & 26.230638 & $27.193616$     & 2  &  19.01    & 1.9               & C & 19.53, 19.63; similar color p-l; both negligible AEN, pm and p\\
014912+422843 & 27.299792 & $42.478624$     & 2  &  17.85   & 2.8              & C & 17.99, 18.82; similar color p-l; both AEN, negligible pm and p\\
015417+433319 & 28.571648 & $43.555321$     & 2  &  18.05   & 2.7                & C & 18.92, 18.28; similar color p-l; both negligible AEN, pm and p\\
022205-234144 & 35.521817 & $-23.69567$     & 2  &  18.99   & 2.1              & C & 18.96, 20.45; similar color p-l; both negligible AEN, pm and p\\
022958+032031 & 37.492401 & $3.341935$      & 2  &  18.02   & 2.1              & C & 18.15, 18.79; similar color p-l; both negligible AEN, pm and p\\
024245-100257 & 40.688737 & $-10.049076$    & 2  &  18.43   & 2.4              & C & 18.73, 19.50; similar color p-l; both negligible AEN, pm and p\\
024950+260651 & 42.459532 & $26.114096$     & 2  &  18.56   & 3.2               & C & 18.81, 20.15; similar color p-l; both negligible AEN, pm and p\\
042022-101932 & 65.092136 & $-10.325513$    & 2  &  18.54   & 3.2              & B & extended+p-l; no Gaia data\\
045048-280957 & 72.701208 & $-28.165922$    & 2  &  18.95   & 5.0               & C & 18.87, 19.07; similar color p-l; both negligible AEN, pm and p\\
051623-043755 & 79.096146 & $-4.631812$     & 2  &  18.20    & 3.0              & C & 18.48, 18.47; p-l; both negligible AEN, pm and p\\
052026-045245 & 80.10733  & $-4.879078$     & 2  &  19.30   & 2.4             & C & 19.57, 19.65; similar color p-l; both negligible AEN, pm and p\\
052902-032948 & 82.260144 & $-3.496646$     & 2  &  19.27   & 1.4              & C & 19.84, 20.37; similar color p-l; both negligible AEN, pm and p\\
061215-193928 & 93.063509 & $-19.657707$    & 2  &  17.49   & 2.2              & C & 18.26, 20.34; similar color p-l (both negligible AEN, pm and p) + red companion;\\
& &  &  &  & & & included in the \citet{delchambre18} Gaia clusters catalogue\\
063019-264851 & 97.580318 & $-26.814116$    & 3  &  18.58   & 3.4              & C & 18.99, 19.05, 19.54; p-l; all have negligible AEN, pm and p; \\
& &  &  &  & & & included in the \citet{delchambre18} Gaia clusters catalogue\\
064505+505755 & 101.269368& $50.965199$     & 2  &  18.84    & 3.0              & C & 19.56, 19.14; similar color p-l; both negligible AEN, pm and p\\
064519+380712 & 101.327789& $38.119957$     & 2  &  17.3      & 2.4              & B & 18.50, 17.63; similar color p-l; both negligible pm and p\\
070249+530654 & 105.704772& $53.114994$     & 2  &  19.05   & 2.5             & C & 19.05, 19.65; p-l; both negligible AEN, pm and p\\
073017+152842 & 112.570702& $15.4782$       & 2  &  18.52   & 2.2              & C & 19.40, 18.78; similar color p-l; both negligible AEN, pm and p\\
081357+103304 & 123.486422& $10.551007$     & 2  &  18.62  & 2.7              & C & 18.97, 18.70; similar color p-l; includes SDSS z=0.799 QSO; \\
& &  &  &  & & & SQLS candidate; both negligible AEN, pm and p\\
081806+524732 & 124.523269& $52.792161$     & 2  &  17.66   & 3.3              & C & 18.96, 17.82; similar color p-l; includes SDSS z=1.793 QSO; \\
& &  &  &  & & & SQLS candidate; both negligible AEN, pm and p\\
085254-014850 & 133.223992& $-1.813836$     & 2  &  18.56  & 3.2             & C & 18.51, 19.94; similar color p-l; both have negligible AEN, pm and p\\
090611-093755 & 136.545112& $-9.632052$     & 2  &  18.76   & 2.8              & C & 18.86, 19.66; similar color p-l; both have negligible AEN, pm and p\\
091724-054200 & 139.348239& $-5.700061$     & 2  &  18.80   & 2.6              & C & 18.87, 19.21; similar color p-l; both have negligible AEN, pm and p\\
092823+213853 & 142.096969& $21.647987$     & 2  &  18.84    & 2.6              & C & 19.05, 19.14; similar color p-l; both have negligible AEN, pm and p\\
094450+243459 & 146.208841& $24.582929$     & 2  &  19.08   & 2.4              & C & 19.84, 20.98; p-l; only one component has Gaia pm and p, negligible values\\
095324+570319 & 148.351564& $57.055364$     & 2  &  18.70   & 2.6              & C & 19.33, 18.90; similar color; includes SDSS z=0.619 QSO; SQLS candidate, \\
& &  &  &  & & & no lensing object; both have negligible AEN, pm and p\\
100406+523132 & 151.025821& $52.525602$     & 2  &  19.45   & 2.3              & C & similar color p-l; no Gaia data\\
100809-044923 & 152.038129& $-4.823158$     & 2  &  18.56     & 2.9              & C & 18.61, 20.14; p-l; only one component has Gaia pm and p, negligible values\\
110928-233315 & 167.366219& $-23.554197$    & 2  &  19.13   & 2.3              & C & 19.47, 20.59; similar color; both have negligible AEN, pm and p\\
111524-030727 & 168.850362& $-3.124282$     & 2  &  18.78   & 2.6              & C & 20.12, 19.06; similar color p-l; both have negligible AEN, pm and p\\
112145+011422 & 170.436445& $1.239436$      & 2  &  19.17   & 1.5             & C & 19.35, 19.81; similar color p-l; includes SDSS z=1.292 QSO; \\
& &  &  &  & & & companion has no Gaia p and pm\\
112456-230507 & 171.233583& $-23.085325$    & 2  &  19.10   & 1.8              & C & 19.48, 19.51; similar color p-l; both have negligible AEN, pm and p\\
113800+073004 & 174.495987& $7.501138$      & 2  &  18.25   & 2.8              & C & 18.39, 19.40; similar color p-l; includes SDSS z=1.209 QSO; SQLS candidate;\\
& &  &  &  & & & no Gaia p and pm\\
115458+185527 & 178.740065& $18.924205$     & 3  &  18.83   & 2.7              & C & 18.88, 20.31; similar color p-l (both have negligible AEN, pm and p) + red \\
& &  &  &  & & & component\\
121410+333703 & 183.540724& $33.617445$     & 3  &  18.91   & 2.5              & B &  19.14, 20.41; similar color p-l (one component has Gaia data, negligible AEN, \\
& &  &  &  & & & pm and p) + red component; includes SDSS z=1.774 QSO, SQLS candidate \\
121410+292445 & 183.541535& $29.412494$     & 2  &  19.47   & 1.5              & C & 19.81; similar color p-l; only one component has Gaia data; negligible AEN, pm \\
& &  &  &  & & & and p\\
121710-025622 & 184.290272& $-2.939367$     & 2  &  19.08   & 1.7              & C & 19.68; similar color p-l; includes SDSS z=1.465 QSO \citep{croom01}; \\
& &  &  &  & & & companion has no Gaia data\\
121756-181837 & 184.481806& $-18.310394$    & 2  &  19.42   & 2.5             & C & 19.44, 20.45; p-l; both have negligible AEN, pm and p\\
130451-102826 & 196.211716& $-10.473908$    & 2  &  19.00    & 2.2             & C & 19.28, 20.14; p-l; both have negligible AEN, pm and p\\
130602+210549 & 196.510055& $21.09696$      & 2  &  18.01   & 2.1              & C & third red component; no Gaia data\\
132202+030933 & 200.508342& $3.159175$      & 2  &  19.33   & 2.7               & C & 19.32; similar color; includes SDSS z=0.961 QSO; SQLS candidate; companion \\
& &  &  &  & & &has no Gaia data \\
135425-094103 & 208.60498 & $-9.684109$     & 2  &  19.05   & 2.1              & C & 19.77; similar color p-l; only one component has Gaia data; negligible values of \\
& &  &  &  & & & AEN, pm and p\\
141855+244107 & 214.731082& $24.685389$     & 2  &  18.88   & 4.5              & C & 19.06, 20.60; similar color p-l; includes SDSS z=0.573 QSO; \\
& &  &  &  & & &  \citep{williams17} candidate; companion has no Gaia data \\
 142816+095443 & 217.065054& $9.911986$      & 2  &  18.63   & 1.8              & C & 18.55, 19.67; p-l; includes SDSS z=1.467 QSO; no lens object; both have \\
 & &  &  &  & & & negligible AEN, p and pm\\
 143125-044338 & 217.854924& $-4.727349$     & 2  &  19.30   & 2.3              & C & 19.30, 20.10; similar color p-l; both have negligible AEN, p and pm\\
 143928-065828 & 219.867271& $-6.974503$     & 2  &  19.17   & 2.3              & C & 19.47, 19.98; similar color p-l; both have negligible AEN, p and pm\\
 144446-163241 & 221.189796& $-16.544779$    & 2  &  18.59   & 2.0              & C & 19.34, 18.95; similar color p-l; both have negligible AEN, p and pm\\
 145939+162155 & 224.914314& $16.365409$     & 2  &  18.67   & 3.5             & C & 18.88, 20.27; similar color p-l; includes SDSS z=1.569 QSO; SQLS candidate; \\
 & &  &  &  & & & both have negligible AEN, p and pm\\
 151545+004328 & 228.936742& $0.724443$      & 2  &  18.96   & 3.5              & C & 19.51, 19.33; similar color p-l; both have negligible AEN, p and pm\\
151546-032231 & 228.941104& $-3.375202$     & 2  &  19.31   & 2.3              & C & 19.53, 20.23; similar color p-l; both have negligible AEN, p and pm\\
\hline
\end{tabular}
\\ 
\label{tab:cand}
\end{minipage}
\end{table*}
\normalsize

\begin{table*}
  \scriptsize
 \centering
 \begin{minipage}{\linewidth}
 \contcaption{}
  \begin{tabular}{@{}lrrclccl@{}}
  \hline 
Name [PS1~J...] & $\alpha$ & $\delta$ & \#Comp & $i$ & Sep. [\arcsec] & Rank & $G$ mags; notes \\ 
 \hline
 152841+393229 & 232.169429& $39.541466$     & 2  &  19.46   & 1.9              & C & 19.61, 20.35; similar color p-l; includes SDSS z=1.215 QSO; both have \\
& &  &  &  & & & negligible AEN, p and pm\\
 153808-192310 & 234.535305& $-19.386104$    & 2  &  19.29   & 2.8              & C & 19.54, 20.43; p-l; both have negligible AEN, p and pm\\
162900-140856 & 247.247099& $-14.148889$    & 2  &  18.55   & 2.4              & C & 19.76, 19.00; similar color; both have negligible AEN, p and pm\\
162903+372433 & 247.260887& $37.409037$     & 2  &  19.05   & 4.3              & C & 19.18, 19.40; similar color p-l; includes SDSS z=0.926 QSO, no lensing object; \\
& &  &  &  & & & both have negligible AEN, p and pm; \citet{williams17} candidate\\
164556+402246 & 251.482344& $40.379443$     & 2  &  19.01   & 2.3              & C & 19.23; similar color p-l; one component has negligible AEN, p and pm, the other\\
& &  &  &  & & &  has no Gaia data\\
165831+141605 & 254.627587& $14.268089$     & 2  &  18.73   & 2.2              & C & 19.11, 19.08; similar color p-l; both have negligible AEN, p and pm\\
 170402+115730 & 256.009503& $11.958322$     & 2  &  18.59   & 2.9              & C & 18.75; similar color p-l; companion has no Gaia data\\
 172406+640711 & 261.027058& $64.119668$     & 2  &  18.16   & 2.4               & C & 18.35, 20.35; similar color p-l; includes SDSS z=1.512 QSO; SQLS candidate;\\
 & &  &  &  & & & both have negligible AEN, p and pm\\
 172751+194436 & 261.960528& $19.743295$     & 2  &  19.32   & 1.9              & C & 20.11; similar color p-l; companion has negligible AEN, p and pm; no Gaia data \\
 & &  &  &  & & & for main component\\
  175526+631504 & 268.857193& $63.251051$     & 2  &  19.28   & 2.2              & C & 19.66, 19.74; similar color p-l; both components have negligible AEN, p and pm\\
 175918+345928 & 269.825014& $34.991208$     & 2  &  19.09   & 2.3              & C & 19.21, 19.64; similar color p-l; both components have negligible AEN, p and pm\\
 183230+534914 & 278.123646& $53.8206$       & 2  &  19.13   & 3.0             & C & 19.58, 20.15; similar color p-l; both components have negligible AEN, p and pm\\
 184624+352002 & 281.599148& $35.333764$     & 2  &  19.33   & 2.4              & C & 19.29, 19.76; similar color p-l; both components have negligible AEN, p and pm\\
 192808+553219 & 292.032689& $55.538539$     & 2  &  18.05   & 2.7              & C & 19.00, 18.32; similar color p-l; both components have negligible AEN, p and pm\\
195243-111715 & 298.179179& $-11.28742$     & 2  &  19.46   & 2.3               & C & 19.70, 20.32; similar color p-l; both components have negligible AEN, p and pm\\
204258-273754 & 310.739743& $-27.631602$    & 2  &  19.15    & 2.3             & C & 19.14, 20.47; similar color p-l; both components have negligible AEN, p and pm\\
205006-225929 & 312.523434& $-22.991253$    & 2  &  18.93   & 2.3              & C & 19.00, 20.22; similar color p-l; both components have negligible AEN, p and pm\\
205143-111444 & 312.931008& $-11.245566$    & 3  &  18.95   & 3.2              & A & 19.66, 19.93, 20.74; p-l sources in quad-like configuration. 2 components have \\
& &  &  &  & & & negligible AEN, p and pm, another has negligible AEN and no other Gaia data; \\
& &  &  &  & & & the final one has no Gaia data\footnote{After the first draft of this work (arXiv:1803.07175v1), this system was independently announced by \citet{delchambre18} as a candidate. Our \texttt{Glafic} modeling of the observed configuration with a SIE$+\gamma$ mass profile results in a perfect fit, but the model is under-constrained because the lensing galaxy is not detected.}\\
212028+280324 & 320.116547& $28.056796$     & 2  &  18.65   & 2.9              & C & 18.78, 19.41; similar color p-l; both components have negligible AEN, p and pm\\
213736+201517 & 324.398524& $20.254669$     & 2  &  19.29   & 1.6              & C & 19.64, 19.66; similar color p-l; both components have negligible AEN, p and pm\\
214132+182621 & 325.382786& $18.439197$     & 2  &  18.97   & 2.4             & C & 18.96, 19.65; similar color p-l; both components have negligible AEN, p and pm\\
214237+255423 & 325.654002& $25.906285$     & 2  &  18.81   & 2.9              & B & 18.76; similar color p-l; only one component has Gaia data, negligible AEN, \\
& &  &  &  & & & p and pm\\
214315+075120 & 325.810482& $7.855534$      & 2  &  18.18   & 2.7              & C & 18.52, 19.00; similar color p-l; both components have negligible AEN, p and pm\\
215034-265214 & 327.643528& $-26.870639$    & 2  &  16.95   & 1.8              & C & includes z=0.115 (lensing?) galaxy \citep{jones09}; no Gaia data\\
215158+111102 & 327.99043 & $11.183861$     & 2  &  18.77   & 2.6              & C & 19.06, 19.74; similar color p-l; includes SDSS z=1.797 QSO; SQLS candidate; \\
& &  &  &  & & & both have negligible AEN, p and pm\\
220943+043217 & 332.428196& $4.538084$      & 2  &  18.18   & 2.9              & C & 18.59, 18.40; similar color p-l; both have negligible AEN, p and pm\\
222108+214518 & 335.283056& $21.754907$     & 2  &  17.34   & 3.5              & C & 17.57; different color p-l; only the main component has Gaia data, negligible \\
& &  &  &  & & & AEN, p and pm\\
230339+345343 & 345.91142 & $34.89518$      & 3  &  18.36   & 7.1                   & C & 18.65, 18.98; similar color p-l (both have negligible AEN, p and pm) + \\
& &  &  &  & & & inner red source\\
231813+025028 & 349.554123& $2.841082$      & 2  &  19.31    & 3.2              & C & 19.59, 19.43; similar color p-l; both have negligible AEN, p and pm\\
232223+375439 & 350.595174& $37.910703$     & 2  &  19.23   & 2.2              & C & 19.87, 20.81; p-l; both have negligible AEN, p and pm\\
232449-122555 & 351.205773& $-12.432025$    & 2  &  18.93   & 1.8              & C & 19.25; p-l; only one component has Gaia data, negligible AEN, p and pm\\
233525+184309 & 353.85286 & $18.71912$      & 2  &  19.31   & 1.9              & C & red components; no Gaia data\\
 \hline
\end{tabular}
\\ 
{\footnotesize Here $\alpha$ and $\delta$ are the right ascension and declination of the candidates in the International Celestial Reference System. ``\#Comp'' refers to the number of components, where we use the number of PS1 sources inside $3\arcsec$ radius, but revise it based on visual inspection, removing spurious sources and counting additional objects which appear to be part of the system. The measured separation (``Sep.'', in arcseconds) is that between the lens candidate point sources or, in case of a quad, the maximum separation between any of the point sources, taken from the PS1 catalogue, or revised as described above. The magnitude is given in $i$-band in the AB system for the brightest resolved component. We quote the \texttt{iMeanPSFMag} measurements from the PS1 catalogue, or the \texttt{Sextractor} \citep{bertin96} \texttt{MAG\_AUTO} in case we had to manually add the brightest component of the system, as described above. We also quote the $Gaia$ $G$-band magnitudes for each system, in the order of increasing separation from the WISE source coordinates. Here ``p-l'' stands for ``point-like'', whereas ``AEN'' (astrometric excess noise), ``p'' (parallax) and ``pm'' (proper motion) are Gaia-based measurements. We list in this table the alphabetic ranking as gravitationally lensed quasars for all systems with an average grade of 1 and above, based on three human graders, as detailed in Section  \ref{sec:search}. We follow the following convention for the alphabetic ranking: A: average grade > 2.5; B: average grade > 1.5; C: average grade $\geq 1$. SQLS refers to the SDSS Quasar Lens Search \citep{inada08,inada10,inada12}. SDSS spectra were searched inside Data Release 14 \citep{abolfathi18}.}
\end{minipage}
\end{table*}
\normalsize

\section{2M1134$-$2103: Imaging data reduction and modeling}
\label{sec:imag} 

\begin{figure}
	\includegraphics[width=\columnwidth]{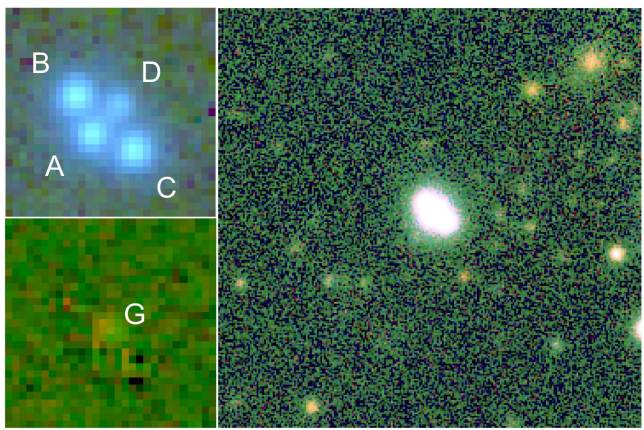}
	\caption{\textit{Upper left:} 
Color composite (VHS-$YJKs$) of 2M1134$-$2103 showing the four lensed
quasar images (A, B, C and D). Image is $10\arcsec$ on the side. 
\textit{Lower left:} The same color composite, after subtracting the four quasar
images with \texttt{hostlens}, shows the presence of a lensing galaxy G. 
\textit{Right:} Color composite
($riz$) using PS1 data shows the immediate environment of the lens
system, which is located at the center. Image is $60\arcsec$ on the side.
All images are oriented such that North is up and East is to the left.
}
    \label{fig:color}
\end{figure}

2M1134$-$2103 consists of four point-like lensed quasar images and a lensing
galaxy (see Fig. \ref{fig:color}).  The lensing galaxy 2M1134$-$2103 can be convincingly
identified in the near-infrared imaging (particularly $Ks$-band) from the VISTA Hemisphere Survey
\citep[hereafter VHS;][see also Fig. \ref{fig:color}]{mcmahon13}. While the relative
astrometry of the quasar images, measured from VST-ATLAS \citep{shanks15}, is
reported in L18, the VST-ATLAS data is not publicly accessible.  Furthermore,
the VST-ATLAS data have better seeing ($0.72\arcsec$) but the PS1 data are
deeper. Therefore, we make use of archival PS1 data in our analysis.  The
processing of the archival PS1 data \citep{flewelling16} is described in
\citep{magnier16a}, and includes removal of the instrumental signature, image
coaddition, as well as photometric and astrometric calibration
\citep{magnier16b}. Here, we model the PS1 $grizy$ and VHS $YJKs$ images
independently of L18.

For our detailed modeling of 2M1134$-$2103 we downloaded from the PS1 and VHS archives $180\arcsec\times180\arcsec$ cutouts around the system in all available filters, large enough to contain stars to model the PSF and to improve the image orientation. We subtracted the sky background from the VHS images using \texttt{SExtractor} \citep{bertin96}, and resampled all images with \texttt{Swarp} \citep{bertin02} to a common orientation. We measured final pixel scales with \texttt{Scamp} \citep{bertin06}.

\begin{figure*}
	\includegraphics[width=\textwidth]{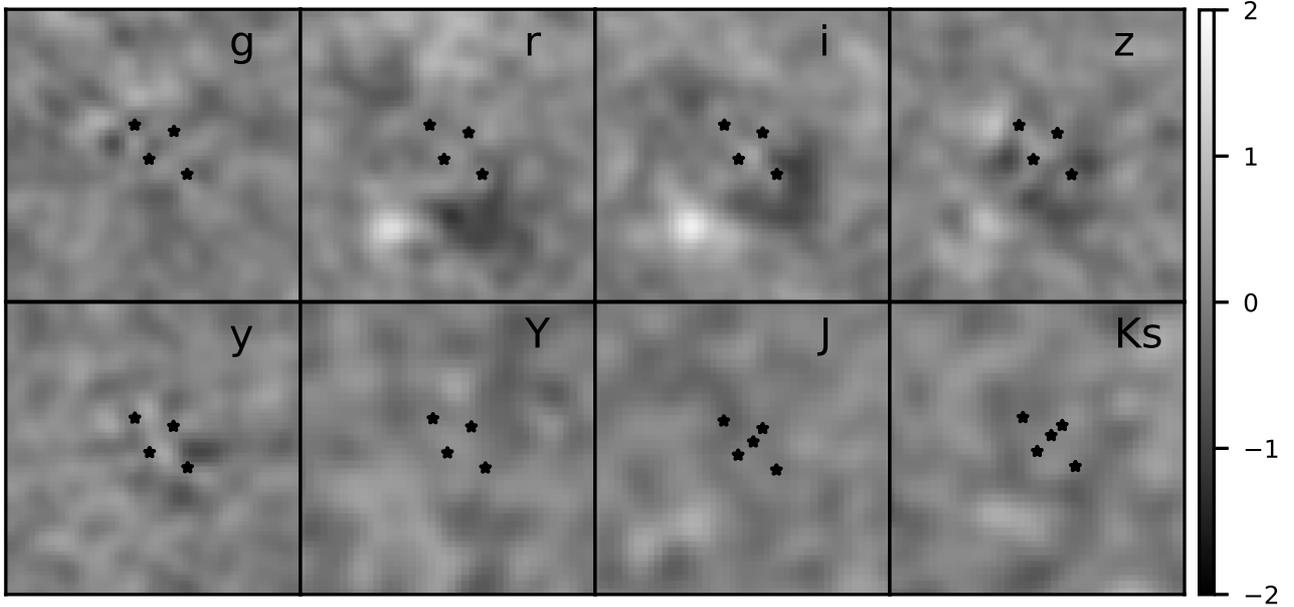}
    \caption{Residuals after morphological modeling of imaging data with \texttt{hostlens}. The size of the cutouts is $15\arcsec\times15\arcsec$. The images were divided by the associated noise maps, then smoothed with a 3-pixel Gaussian, to enhance structure. The positions of the components that were modeled in each band (A, B, C, D, as well as G in $JKs$) are marked with star symbols. Object GX to the south-east of the lens, conspicuous in the $r$ and $i$ bands, is left unmodeled (see Section~\ref{sec:lens}) .} 
    \label{fig:hostlens}
\end{figure*}

We model the system with \texttt{hostlens} \citep{rusu16}. \texttt{Hostlens} models an arbitrary number of point-like and extended sources using a common point-spread function (PSF), either specified by the user from nearby stars, or fitted to the data as a sum of two concentric Moffat \citep{moffat69} profiles. We find that modeling the quasar images using nearby stars as PSFs results in significant residuals, which could affect the image flux measurements and the characterization of the lensing galaxy. We therefore model the data using an analytical PSF fitted to the data. To remove residuals still remaining at the centers of the three bright quasar images in the $rizYJKs$ bands, we use the PSF reconstruction technique described in \citet{chen16}, with the best-fit analytical PSF as a starting point. This technique reconstructs the PSF iteratively, on a grid of pixels, under the assumption that the PSF does not vary across the quasar images. The remaining residuals at the location of the quasar images are small, as can be seen in Fig.~\ref{fig:hostlens}.

In the PS1 data we do not detect any sign of the lensing galaxy, which however stands out in the VHS $J$ and $Ks$ images. We model its light profile in these bands simultaneously with the quasar images, using a de Vaucouleurs \citep{devaucouleurs48} profile commonly used for early-type lensing galaxies. A circular profile fits the emission from the lensing galaxy well, without leaving noticeable residuals. Using a Markov Chain Monte Carlo (MCMC) approach, we find that the lensing galaxy flux is highly degenerate with the effective radius of the de Vaucouleurs profile, and is therefore unreliable. 

In order to perform gravitational lens modeling of 2M1134$-$2103 we need to estimate reliable relative astrometry for the quasar images and the lensing galaxy. For the three brightest quasar images we take the mean and scatter between the measured relative astrometry in different filters (excluding $g$-band, where the seeing is significantly larger, see Table~\ref{tab:data}), whereas for the lensing galaxy and the faint counter-image D,  we only use the $J$ and $Ks$ filters. Indeed, the separation between the brighter images (A, B and C) and the fainter counter-image (D) decreases slightly with increasing wavelength in the PS1 images, because of the progressively increasing flux contribution from the red lensing galaxy. We report our measured astrometry and photometry in Table \ref{tab:data}. Our astrometry is consistent with the one presented in L18 within our $2\sigma$ uncertainties.

\begin{table*}
 \centering
 \begin{minipage}{\linewidth}
  \caption{Relative astrometry and absolute photometry of 2M1134$-$2103}
  \begin{tabular}{@{}lllllccl@{}}
  \hline 
Filter (lim. mag) & A & B & C & D & G & GX & Seeing [$\arcsec$] \\ 
 \hline
all (x-axis) & $0.000\pm0.000$ & $-0.733\pm0.005$ & $1.944\pm0.006$ & $1.262\pm0.014$ & $0.74\pm0.04$ & $-2.50\pm0.05$ &  \\
all (y-axis) & $0.000\pm0.000$ & $1.757\pm0.006$ & $-0.776\pm0.006$ & $1.350\pm0.020$ & $0.75\pm0.10$ & $-3.32\pm0.05$ &  \\
 \hline
$g$ (24.2) & $17.08\pm0.005$ & $17.37\pm0.005$ & $17.26\pm0.005$ & $18.90\pm0.014$ & $-$ & $-$ & $1.70$ \\
$r$ (24.5) & $16.85\pm0.005$ & $17.06\pm0.005$ & $17.00\pm0.005$ & $18.67\pm0.005$ & $-$ & [$23.37\pm0.10$] & $1.20$ \\
$i$ (24.5) & $16.81\pm0.005$ & $16.88\pm0.005$ & $16.83\pm0.005$ & $18.46\pm0.005$ & $-$ & [$21.75\pm0.08$] & $1.20$ \\
$z$ (23.6) & $16.87\pm0.005$ & $16.90\pm0.005$ & $16.87\pm0.005$ & $18.49\pm0.006$ & $-$ & $-$ & $1.10$ \\
$y$ (22.6) & $16.79\pm0.04$ & $16.72\pm0.03$ & $16.70\pm0.03$ & $18.29\pm0.04$ & $-$ & $-$ & $1.00$ \\
$Y$ (22.2) & $16.08\pm0.008$ & $15.98\pm0.007$ & $16.02\pm0.007$ & $17.57\pm0.016$ & $-$ & $-$ & $0.85$ \\
$J$ (21.4) & $15.92\pm0.005$ & $15.81\pm0.005$ & $15.83\pm0.005$ & $17.35\pm0.012$ & [$19.05\pm0.12$] & $-$ & $0.85$ \\
$Ks$ (20.1) & $15.34\pm0.006$ & $15.13\pm0.006$ & $15.19\pm0.009$ & $16.81\pm0.027$ & [$17.33\pm0.09$] & $-$ & $0.85$ \\
\hline
\end{tabular}
\\ 
{\footnotesize Relative astrometry is determined by using information from multiple filters (See Section~\ref{sec:imag}). The units are arcseconds and the sign convention is positive from E to W (x-axis) and from S to N (y-axis). The ICRS position of image A in the PS1 catalogue is (J2000.0) 11:34:40.588 $-$21:03:23.06. Magnitudes are in the AB ($grizy$) and Vega ($YJKs$) systems, and are corrected for Galactic extinction following \citet{schlafly11}. The $1\sigma$ limiting magnitudes are computed in $2\arcsec$-radius blank sky apertures around the system. The errors on magnitudes are those from MCMC, with the minimum uncertainty boosted to 0.005 mag, and do not include zeropoint or PSF uncertainties. The magnitudes of G and GX (see Section \ref{sec:lens}) should be considered unreliable, as in order for the fit to converge, the effective radius was fixed to $<1$ pixel.}
\label{tab:data}
\end{minipage}
\end{table*}

\section{2M1134$-$2103: Keck Spectroscopy}
\label{sec:spec} 

The 2M1134$-$2103 lens system was observed with the Echellette
Spectrograph and Imager \citep[ESI;][]{ESI} on the night of 2017 Nov
18 UT (program number 2017B\_U110).  The observations utilized a slit with a width of
1$^{\prime\prime}$ and the cross-dispersed echellette mode of the
spectrograph, which provides a constant dispersion of roughly
11.5~km~sec$^{-1}$~pix$^{-1}$ over a wavelength range of approximately
3900 to 11000~\AA. Here, we follow the nomenclature of L18. Two slit
position angles were used, one oriented at $+46.7^\circ$ (N through E)
in order to go across lensed images B and C (henceforth the ``BC slit'')
and one oriented at $-42.3^\circ$ to cover images A and D (the ``AD
slit'').  We obtained three 600~s exposures through the BC slit and four
600~s exposures through the AD slit.

\begin{figure*}
\includegraphics[width=0.8\textwidth]{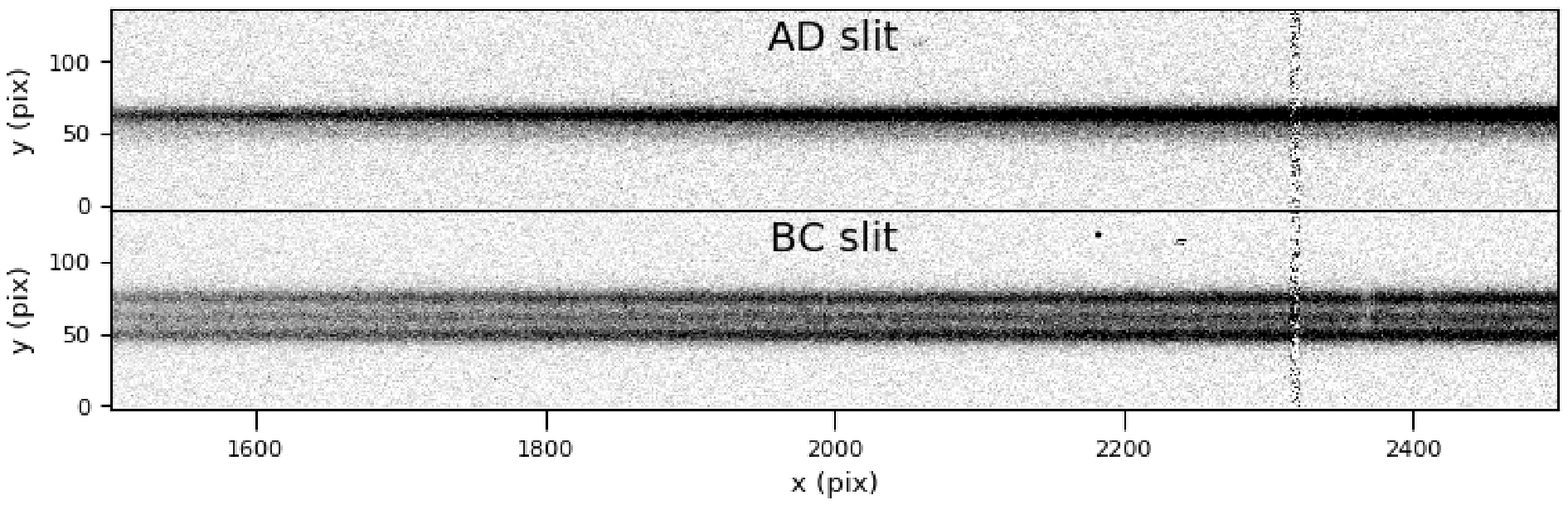}
\caption{Examples of the calibrated and sky-subtracted spectra obtained with
Keck/ESI of 2M1134$-$2103.  Data are from the AD slit (top) and the BC slit (bottom), with
spectra showing a portion of the fifth of ten spectral orders recorded
by the spectrograph.} 
\label{fig:2dspec}
\end{figure*}

We calibrated the data using a custom pipeline written in Python.  The
pipeline does a flat-field correction, rectifies the two dimensional
spectra, does the wavelength calibration using both arc lamp and night
sky emission lines, and subtracts the sky emission.  The calibrated
data for both slits are shown in Fig.~\ref{fig:2dspec}.  In the AD
slit, the two-dimensional spectra show one bright trace that is
heavily blended with a much fainter trace, while the BC slit shows
three clearly separated traces.  We identify the three traces in the
BC slit with components B, A$+$D$+$G, and C in the imaging data. Based on the
imaging, we expect that the emission from image A completely dominates
the central trace.  

\begin{figure*}
\includegraphics[width=\linewidth]{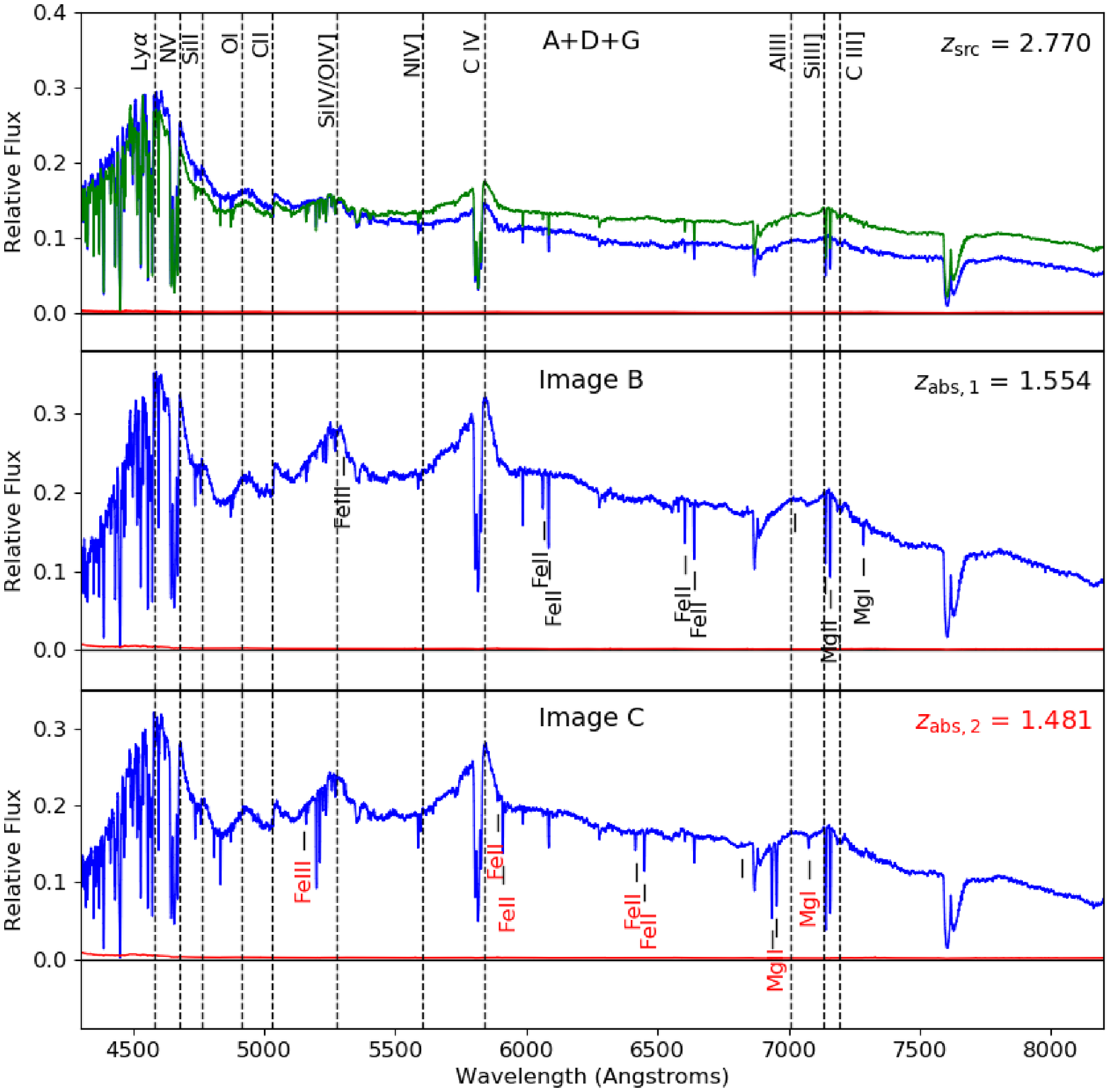}
\caption{ESI spectra of 2M1134$-$2103. Combined light from lensed images A and D, plus any emission from the lensing galaxy is visible in the spectrum extracted from the AD slit (green) as well as that from the BC slit (dark blue) in the top panel. Spectra of images B (middle) and C (bottom) are extracted from the BC slit. The rms noise of each spectrum is plotted in red. Identified emission and absorption systems are labelled (see further explanation in Section~\ref{sec:spec}). The wide absorption doublet at $\sim 7650$ \AA\ is telluric. The spectra were smoothed using a 3-pixel boxcar with inverse-variance weighting.} 
\label{fig:3spec}
\end{figure*}

We extracted one-dimensional spectra from the exposures on both slits
using a second Python pipeline that extracts the spectra from each
spectral order, applies a response correction based on observations of a
spectrophotometric standard, in this case Feige~110, and finally
combines the data from each of the 10 spectral orders into one final
spectrum.  For the AD slit, we only extracted one aperture that we
identify with a blend of A, D, and G, while for the BC slit we extracted
separate apertures corresponding to B, A$+$D$+$G, and C.
Note that the AD slit may very well contain significant scattered light
from images B and C.  The extracted spectra are shown in
Fig.~\ref{fig:3spec} and all show clear broad emission lines that,
furthermore, are indicative of quasars at a redshift of $z_{\rm src}
\sim 2.77$. Thus, the ESI spectra are fully consistent with the
interpretation of 2M1134$-$2103 as a quad lensed quasar. An exact value
of the source redshift is difficult to obtain due to the fact that the
peaks of the lines used for redshift determination are affected by
absorption systems \citep[Fig.~\ref{fig:3spec}; also e.g.,][]{lee18}.
The measured redshift is smaller than the $z\sim3.5$ estimate in L18,
based on PS1 colors.

In addition to the broad emission lines, all of the spectra show a
number of absorption lines. In the range 5000--7500\AA, these
correspond to absorption features of Fe and Mg, and are consistent
with two separate absorption systems at $z_{\rm abs,1} = 1.554$ and
$z_{\rm abs,2} = 1.481$. The first system has stronger lines in the
A$+$D$+$G and B spectra, while the second is stronger in the image C
spectrum.  Although it is possible that these systems may be 
associated with the primary lensing galaxy, the
narrowness of the lines makes this interpretation unlikely. A
much stronger indication of the lensing galaxy would be the detection
of stellar absorption lines, such as the Ca{\sc II} H and K lines, with widths
consistent with the velocity dispersions of $>$100~km s$^{-1}$
expected for a massive lensing galaxy. If these corresponded to the redshifts of the absorption features mentioned above, they would be observed at wavelengths longer than the ones plotted in Fig.~\ref{fig:3spec}, where we have extracted robust spectra.

\section{2M1134$-$2103: Gravitational lens modeling} 
\label{sec:lens} 

We perform gravitational lens mass modeling of 2M1134$-$2103 with \texttt{glafic} \citep{oguri10a}, using the observed relative positions of the quasar images and the lensing galaxy as constraints. We do not impose constraints based on the flux ratios, as these might be affected by microlensing, extinction, and intrinsic variability \citep[e.g.,][]{yonehara08}. However, we analyze the observed flux ratios under the assumptions that they are dominated by extinction, in Section \ref{sec:redshift}. 

We start with the same mass model used in L18, a singular isothermal sphere with external shear (SIS$+\gamma$). This model has $\chi^2/\mathrm{d.o.f.}=7.5/3$ (where $\mathrm{d.o.f.}$ stands for degrees of freedom), most of which is due to the difference between the measured and predicted position of image D relative to the lens G. We recover the results of L18, in particular that an unusually large shear of $\sim0.34$ at 44 deg W of N is required to fit this system. 

Secondly, we fit a model which allows for mass ellipticity, SIE$+\gamma$. Indeed, quads have enough constraints to disentangle internal and external sources of shear, and our fit shows a dramatic improvement to $\chi^2/\mathrm{d.o.f.}=0.1/1$. This model requires a shear of $\sim0.39$, slightly larger than before, and a mass axis ratio of $0.80^{+0.10}_{-0.18}$, with the long axis at $37^{+5}_{-13}$ deg E of N, almost perpendicular to that of the shear. While our imaging data does not have sufficient resolution to fit an elliptical light profile to the lensing galaxy, studies of quads show that the mass and light profiles of lensing galaxies are typically aligned \citep[e.g.,][]{keeton98,sluse12}. We note that \citet{rusu18} modeled a different quad, GraL~J1817+2729, in a cross-like configuration, and showed that while an SIE$+\gamma$ model required large shear and large ellipticity, with the long axis perpendicular to the shear, similar to the present case, a Sersic+SIS$+\gamma$ model, where the Sersic component stands for the baryonic matter in the disk of the lensing galaxy, significantly diminishes the required shear, and changes its orientation. We attempted to fit such as model here, but it behaves equivalently to our SIE$+\gamma$ model, requiring similar orientation and large shear. It appears that the highly stretched, diamond-like configuration, cannot be explained by internal sources of shear.

As both the SIS$+\gamma$ and SIE$+\gamma$ models are consistent in their requirement of large external shear, we look for potential sources of shear from the surrounding environment. In Fig.~\ref{fig:color} we display a $60\arcsec\times60\arcsec$ color composite image around 2M1134$-$2103, which clearly shows a group of red galaxies in the upper right corner, the brightest of which is a $i=19.32$ galaxy located at $\alpha=173.6620$, $\delta=-21.0502$, $30\arcsec$ from the quad, in the direction of 45 deg W of N.  The PS1 and VHS colors of this galaxy imply a photometric redshift of $0.70\pm0.09$, estimated with \texttt{BPZ} \citep{benitez00}. The existence of the galaxy group at this location implies that it is responsible for part of the measured shear. However this is unlikely to be the complete picture, as an SIS profile at the location of this galaxy would require a very large velocity dispersion $\gtrsim1100$ km/s to produce the measured shear, depending on the redshift of the lensing galaxy in 2M1134$-$2103. 

Fig.~\ref{fig:hostlens} reveals another clue, closer to 2M1134$-$2103. After subtracting the quasar images, an additional component is detected in filters $r$ and $i$, $4.16\arcsec$ from image A, also in the direction of the shear, towards south-east. It is unclear whether this new component, which we name GX, is a galaxy or a star, as it is too faint ($i\sim21.75$) to constrain its morphology. Under the assumption that it is a galaxy, its colors suggest a redshift lower than the one of the lensing galaxy, which is only detected in the near-infrared VHS filters. We incorporate GX into a third lensing model, in order to estimate its effect on the external shear. As we do not know the redshifts for either G or GX, we consider the simplest case in which G and GX are modeled as SIS of equal strength, at the same redshift. This model is expected to be an upper limit to the contribution of GX to the lensing configuration, as GX is likely a lower redshift, low mass galaxy. We obtain a good fit with $\chi^2/\mathrm{d.o.f.}=2.3/3$, and a residual external shear of $\sim0.19$, oriented as before. In this model, the two lenses are located $\sim4$ Einstein radii apart, in units of the Einstein radius of G. Our model shows that GX can explain a significant fraction of the shear we measured in our initial SIS$+\gamma$ model. We expect that in reality most of the measured shear is an interplay between the effects of GX and the nearby group. In our model incorporating GX, the nearby group would still require a velocity dispersion of $\sim800$ km/s to account for the remaining shear, which would imply $\gtrsim50$ group/cluster members \citep[e.g.,][]{berlind06}. While we do not see more than $\sim 3$ possible galaxy members in the PS1 image, this is not an argument against the existence of this structure, as the PS1 images are shallow. Indeed, PS1 images of the system RX~J0911+05 reveal only $\sim5$ galaxies part of a spectroscopically confirmed cluster with at least 24 members at a similar redshift of $z=0.769$, with velocity dispersion $\sim800$ km/s \citep{kneib00}, giving rise to a very large shear $\gtrsim0.3$ \citep{sluse12}. 

We note that another lensed system with a remarkably similar diamond-like configuration has recently been discovered \citep{bettoni19} close to a galaxy cluster, also with a large measured shear of 0.31 and a nearby galaxy in the direction of the shear. Finally, we note that highly-sheared quadruple lens systems are not unexpected, and are a consequence of the tendency of elliptical galaxies, which constitute most of the lensing galaxies, to reside in overdense regions with high shear \citep[e.g.,][]{holder03}. We conclude that the large shear values we measure do not, therefore, point out to a problem with our mass models. 

In the analysis above we did not assume particular values of source and lens redshifts, except when we estimated the velocity dispersion of the galaxy group at $z\sim0.7$. The flux ratios are also insensitive to the choice of redshifts, however the estimated time delays depend on them. To estimate the time delays, which are of interest to cosmography studies \citep[e.g,][]{bonvin17}, we use the source quasar redshift $z_s=2.77$ measured from spectroscopy, and the lens redshift limits we infer below in Section \ref{sec:redshift}, $z_l\sim0.45-1.5$. For $z_l\sim0.45$ and the SIS$+\gamma$ model, the estimated time delays are $\Delta \mathrm{CB}\sim7$ days, $\Delta \mathrm{CA}\sim30$ days and $\Delta \mathrm{CD}\sim55$ days. The order of the image time arrival is the same in all three models, with image C leading. We summarize  the main parameters of the mass models we employed in Table~\ref{tab:glafic}, along with the corresponding time delays.

\begin{table*}
 \centering
 \begin{minipage}{\linewidth}
  \caption{Summary of the best-fit parameter values of the lensing mass models, and the predicted time delays}
  \begin{tabular}{@{}rrrrrrrrrr@{}}
  \hline 
Model & $z$ & $\sigma$ [km/s] & $e$ & $\theta_e$ & $\gamma$ & $\theta_\gamma$ & $\Delta\mathrm{CA}$ & $\Delta\mathrm{CB}$ & $\Delta\mathrm{CD}$ \\ 
 \hline
SIS$+\gamma$ & 0.45 & 243.0 & $-$ & $-$ & 0.34 & 43.6 & 30.5 & 6.8 & 54.8 \\
SIS$+\gamma$ & 1.50 & 384.1 & $-$ & $-$ & 0.34 & $43.6$ & 196.6 & 43.9 & 353.1 \\
SIE$+\gamma$ & 0.45 & 242.1 & 0.33 & $-39.9$ & 0.39 & 45.2 & 24.7 & 6.3 & 43.6 \\
SIE$+\gamma$ & 1.50 & 382.6 & 0.33 & $-39.9$ & 0.39 & 45.2 & 159.0 & 40.3 & 281.0 \\
2SIS$+\gamma$ & 0.45 & 233.3 & $-$ & $-$ & 0.19 & 45.8 & 33.9 & 7.5 & 44.3 \\
2SIS$+\gamma$ & 1.50 & 363.4 & $-$ & $-$ & 0.19 & 45.8 & 225.9 & 49.3 & 263.4 \\
\hline
\end{tabular}
\\ 
{\footnotesize Here $z$ is the lens redshift, $\sigma$ is the lens velocity dispersion, $e$ and $\gamma$ are the lens ellipticity and shear, respectively, and $\theta_e$ and $\theta_\gamma$ are their orientations (W of N). The time delays (last column) are in units of days.}
\label{tab:glafic}
\end{minipage}
\end{table*}

\subsection{Flux ratio analysis and the lens redshift}
\label{sec:redshift} 

\begin{figure}
	\includegraphics[width=\columnwidth]{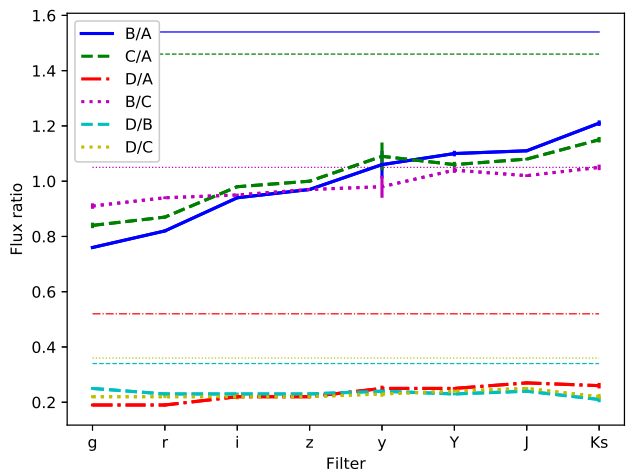}
    \caption{Measured and model-predicted flux ratios of the four quasar images. Thick lines connect observed flux ratios in all available filters, and thin horizontal lines of corresponding colors mark the flux ratios predicted by the SIS$+\gamma$ lens mass model.}
    \label{fig:fluxratio}
\end{figure}

We show the measured image flux ratios in Fig.~\ref{fig:fluxratio}, based on Table \ref{tab:data}. At least three of the six ratios show a clear dependence on wavelength. Interpreted as due to extinction, these ratios imply that A is the least reddened image, in agreement with image D being closest to the lensing galaxy, and the major axis of the lensing galaxy being oriented towards B and C (with B more reddened then C), according to the SIE$+\gamma$ model. We also show in Fig.~\ref{fig:fluxratio} the predicted flux ratios given by the SIS$+\gamma$ model. The predicted SIE$+\gamma$ fluxes are very similar, within $\sim10\%$, but the G$+$GX$+\gamma$ model predicts a demagnified image A, about as bright as D. All models predict image B to be the brightest, as observed in the VHS data. The predicted B/C is invariant across all three models, as the shear is almost perpendicular to the direction of these two images. In fact, this ratio is the only one which matches the observations, in the reddest filter.

Flux ratios of quasar images have been used in the past to study the extinction properties of lensing galaxies \citep[e.g.,][]{falco99} as well as to infer lens redshifts \citep[e.g.,][]{jean98}. Here, we use them to infer the lens redshift $z_l$ as well as the de-reddened flux ratios (relative magnifications) $M_i$, where $i$ refers to each of the six image pairs, independent of the chosen mass model. Following \citet{falco99}, we optimize these parameters as well as the differential extinctions $E_i$ and the shape of the extinction curve $R$ by minimizing

\begin{equation}
\chi^2=\sum_{j=1}^{N_\lambda}\sum_{i=1}^{N_\mathrm{imag}}{\frac{\left[m^r_i(\lambda_j)-m^b_i(\lambda_j) - 2.5\log M_i - E_iR\left(\frac{\lambda_j}{1+z_l}\right)      \right]^2}{\sigma^{b,2}_{ij}+\sigma^{r,2}_{ij}}}
\end{equation}

\noindent where $j$ is the filter index, superscripts $b$ and $r$ refer to the blue and red images in each pair, respectively, and $\sigma_{ij}$ is the magnitude measurement uncertainty. We use the central wavelength of each filter, and the \citet{cardelli89} extinction function implemented in the code \texttt{extinction}\footnote{\url{http://extinction.readthedocs.io/en/latest/}}. We perform the minimization using the Nelder-Mead \citep{nelder65} method implemented in \texttt{Scipy} \citep{oliphant07}, starting from random positions in the parameter space and further exploring around the solution with \texttt{emcee} \citep{foreman13}, to ensure that we have found the global solution. 

We find the best-fit solution ($\chi^2/\mathrm{d.o.f.}=213.4/16$)\footnote{With four images, thus three independent flux ratios in each band, and with eight bands, we have 24 constraints. As parameters, we have the redshift, the extinction curve parameter, three independent extinctions and 3 independent magnifications, thus eight parameters, resulting in 16 degrees of freedom.} with $z_l\sim0.45$, $R\sim2.5$ slightly smaller than the Galactic extinction curve with $R_V=3.1$, small $E_i\lesssim0.1$ consistent with the results in \citet{falco99}, and flux ratios B/A $=1.28$, C/A $=1.20$, D/A $=0.30$. These parameter values are robust if we remove from the fit all image pairs containing D (new fit $\chi^2/\mathrm{d.o.f.}=145.8/10$), in case our decomposition of G and D is problematic due to the low image resolution. They are also robust to the choice of the extinction function. Except for B/C, which matches the prediction of the mass models, the flux ratios are smaller than predicted. The quality of the fits is statistically poor, although such large $\chi^2$ values are found by e.g., \citet{falco99} in other lensing systems as well. In our analysis, we have ignored any contribution from microlensing and quasar intrinsic variability, which can also affect flux ratios chromatically \citep[e.g.,][]{yonehara08}.

\begin{figure*}
	\includegraphics[width=\linewidth]{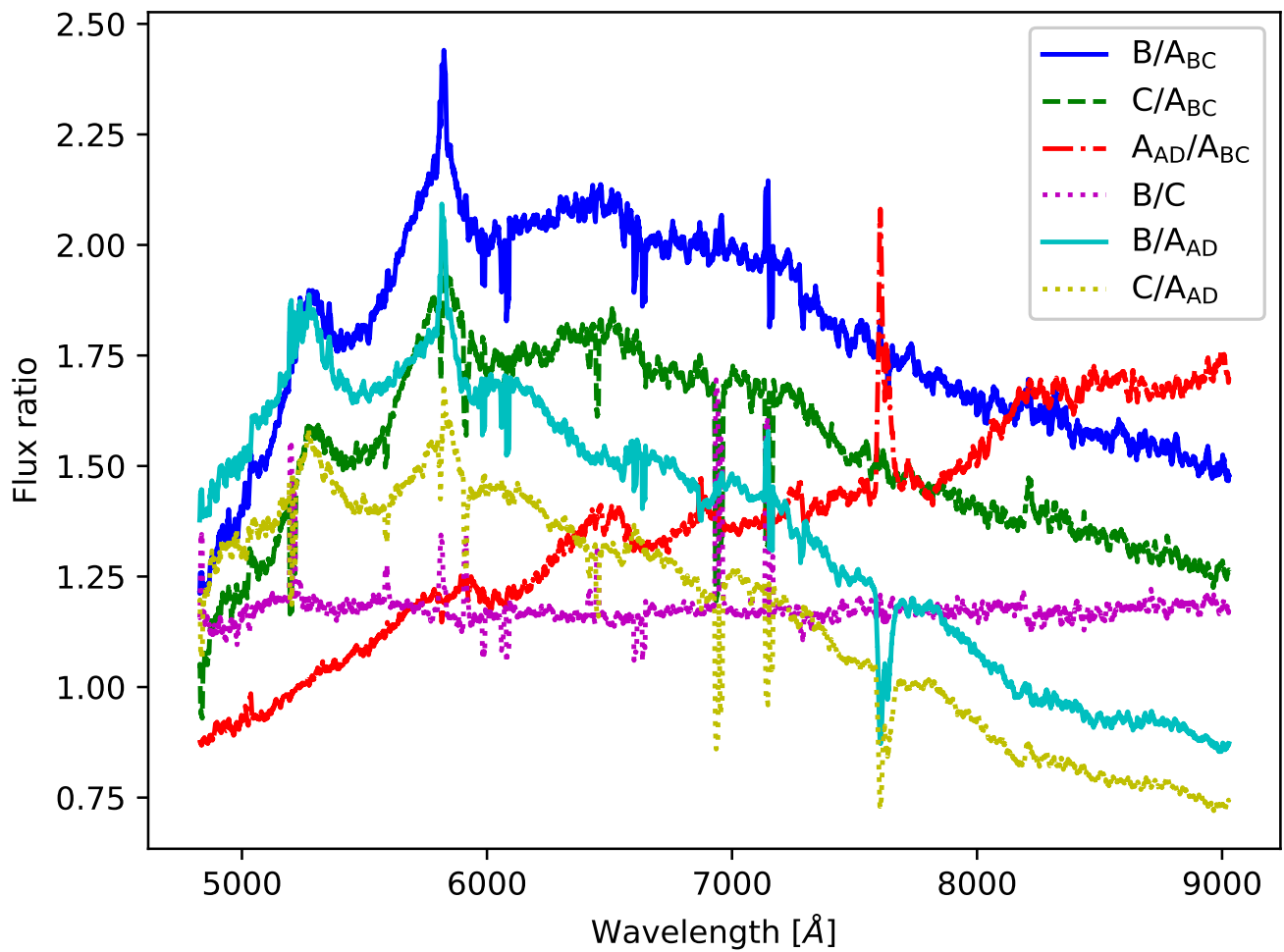}
    \caption{Measured spectral flux ratios in the region 4830 \AA - 9030 \AA, where the measurements are robust. The colors and line styles correspond to those in Fig. \ref{fig:fluxratio}, except that some flux ratios have been inverted and photometric ratios involving image D have been replaced with spectroscopic ratios of $A_{AD}$. The spikes correspond to the intrinsic or atmospheric absorption lines in the original spectra.}
    \label{fig:specratio}
\end{figure*}

We can look for signs of microlensing by plotting the quasar image spectral ratios. While the overall shape of these ratios is sensitive to observational effects such as sub-optimal slit placement and differential refraction, these (as well as differential extinction) should affect both continua and emission lines equally. On the other hand, microlensing is dependent on the size of the source, such that the continuum emission, which originates from a more compact region than the broad emission lines, should be preferentially microlensed. Fig. \ref{fig:specratio} clearly shows that, when dividing the fluxes of B and C to those of A$_\mathrm{BC}$ (i.e. the A+D+G signal, dominated by A, and extracted from the BC slit) and of A$_\mathrm{AD}$, there is a large jump in the flux ratios at the locations of the SiV/OIV] ($\sim 5270$\AA) and CIV ($\sim 5800$\AA) broad line regions, compared to the surrounding continuum. On the other hand, B/C is relatively flat over the entire plotted range, which means that microlensing affects image A+D (the saddle points of the time arrival surface) but not B and C. A direct comparison of the photometric flux ratios in Fig. \ref{fig:fluxratio} with the spectroscopic ratios in Fig. \ref{fig:specratio} is not possible due to the fact that the spectra are affected by slit losses. Indeed, this can be seen from the monotonic variation in A$_\mathrm{AD}$/A$_\mathrm{BC}$, which we attribute to the fact that ESI does not use an atmospheric dispersion corrector, thus resulting in flux losses from differential refraction, particularly between the orthogonally placed AD and BC slits. Also, the datasets are not concurrent, and are therefore prone to time-varying microlensing and intrinsic variability effects.

As discussed above, microlensing in particular may affect the inferred lens redshift. We note that due to the low image resolution, proximity to image D, and morphological compactness which may affect the extracted photometry, we could not obtain a robust photometric redshift for this galaxy. Looser but more robust redshift constraints can be set by using the observed image separation and the estimated magnitude of the lens in the filter in which it is brightest. On the one hand, the image separation gives the lens velocity dispersion as a function of redshift; on the other, assuming an early-type spectral template, the measured magnitude can be converted into a rest-frame absolute magnitude as a function of redshift\footnote{We use the \texttt{mag2mag} routine from \citet{auger09}, available at \url{https://github.com/tcollett/LensPop/tree/master/stellarpop/}}, and then into a velocity dispersion \citep{faber76}. We find a lower limit of $z_l\sim0.5$, below which the two velocity dispersion estimates disagree, and an upper limit of $z_l\sim1.5$, above which the lens velocity dispersion is $\sim400$ km/s, a value above which the galaxy velocity dispersion function is vanishingly small \citep{sheth03}. The lower limit is close to the value inferred from our flux ratio analysis, and the upper one is consistent with the redshift of the narrow absorption systems identified in Section \ref{sec:spec}; it is also above the L18 estimate of $z_l\sim1$.

\section{Conclusions and Future Work}
\label{sec:conc} 
We have carried out a systematic search for gravitationally lensed quasars in PS1, based on visual examination of cutouts around the AGN source catalogue of \citet{secrest15}, and aided by astrometric quantities measured by $Gaia$. We present our sample of 91 promising candidates, not found in the available literature, in Table~\ref{tab:cand}, in order to enable follow-up observations by the interested community. We expect that the main source of contaminants are quasar pairs, which we see as a byproduct of our work, and to a lesser extent, quasar + star pairs. Our best estimate of the purity of our sample, in terms of lensed doubles and quasar pairs contaminated by quasar + star pairs, is $\sim70\%$.

As part of our search, we have independently discovered six known quads, including 2M1134$-$2103. We present, for the first time, spectroscopy of this system, confirming it as a lensed quasar with source redshift $z\sim2.77$. We identify absorption systems at $z\sim1.5$, in three of the resolved quasar images, but we find these to be too narrow to attribute to the lensing galaxy. The image flux ratios show a monotonic dependence on wavelength, which we use to obtain a rough estimate of the lens redshift, under the assumption that the dependence is caused by extinction. The spectral flux ratios show evidence of microlensing in the combined emission from images A and D. 

Our mass modeling confirms that 2M1134$-$2103 is affected by large shear, for which we identify two potential sources: a group of galaxies at $z\sim0.7$, $30\arcsec$ from the lens, and another faint companion $\sim4\arcsec$ away. Future multi-object spectroscopy is required to determine whether these are part of a larger cluster, or physically associated with the lens. The large image separation, brightness and estimated time delays ranging from several days to several months, depending on the lens redshift, make this a valuable system to use for cosmography \citep[e.g.,][]{bonvin17}, provided that the environment can be characterized with future, deep imaging and spectroscopy \citep[e.g.,][]{sluse17,rusu17,wilson16}. High resolution \textit{Hubble Space Telescope} or adaptive optics imaging is necessary to constrain the morphology of the lensing galaxy, and to further constrain the mass models using the expected extended emission from the underlying host galaxy \citep[e.g.,][]{chen16,wong17}. 

\section*{Acknowledgements}

The authors are indebted to the anonymous referee for thoroughly checking the sample of candidates, resulting in a significant improvement of our selection, and for suggesting the use of $Gaia$. We also wish to thank Paul L. Schechter for his ``matchmaking'' which has made this paper possible. This research made use of \texttt{Astropy}, a community-developed core Python package for Astronomy \citep{astropy13}. Plots were produced with \texttt{Matplotlib} \citep{hunter07}. This work was supported by World Premier International Research Center Initiative (WPI Initiative), MEXT, Japan.  C.D.F. acknowledges support from the National Science Foundation under Grant No. AST-1715611.

The Pan-STARRS1 Surveys (PS1) and the PS1 public science archive have been made possible through contributions by the Institute for Astronomy, the University of Hawaii, the Pan-STARRS Project Office, the Max-Planck Society and its participating institutes, the Max Planck Institute for Astronomy, Heidelberg and the Max Planck Institute for Extraterrestrial Physics, Garching, The Johns Hopkins University, Durham University, the University of Edinburgh, the Queen's University Belfast, the Harvard-Smithsonian Center for Astrophysics, the Las Cumbres Observatory Global Telescope Network Incorporated, the National Central University of Taiwan, the Space Telescope Science Institute, the National Aeronautics and Space Administration under Grant No. NNX08AR22G issued through the Planetary Science Division of the NASA Science Mission Directorate, the National Science Foundation Grant No. AST-1238877, the University of Maryland, Eotvos Lorand University (ELTE), the Los Alamos National Laboratory, and the Gordon and Betty Moore Foundation.

This work has made use of data from the European Space Agency (ESA)
mission {\it Gaia} (\url{https://www.cosmos.esa.int/gaia}), processed by
the {\it Gaia} Data Processing and Analysis Consortium (DPAC,
\url{https://www.cosmos.esa.int/web/gaia/dpac/consortium}). Funding
for the DPAC has been provided by national institutions, in particular
the institutions participating in the {\it Gaia} Multilateral Agreement.

Based on observations obtained as part of the VISTA Hemisphere Survey, ESO Progam, 179.A-2010 (PI: McMahon). Based on data obtained from the ESO Science Archive Facility under request number
cerusu312487. 

Some of the data presented herein were obtained at the W. M. Keck Observatory, which is operated as a scientific partnership among the California Institute of Technology, the University of California and the National Aeronautics and Space Administration. The Observatory was made possible by the generous financial support of the W. M. Keck Foundation.

The authors recognize and acknowledge the very significant cultural role and reverence that the summit of Mauna Kea has always had within the indigenous Hawaiian community. We are most fortunate to have the opportunity to conduct observations from this superb mountain.


\begin{thebibliography}{99}
\bibitem[\protect\citeauthoryear{Abolfathi et al.}{2018}]{abolfathi18} Abolfathi B., et al., 2018, ApJS, 235, 42
\bibitem[\protect\citeauthoryear{Agnello et al.}{2015}]{agnello15} Agnello A., Kelly B.~C., Treu T., Marshall P.~J., 2015, MNRAS, 448, 1446 
\bibitem[\protect\citeauthoryear{Agnello et al.}{2018}]{agnello18b} Agnello A., et al., 2018, MNRAS, 475, 2086 
\bibitem[\protect\citeauthoryear{Agnello et al.}{2018a}]{agnello18a} Agnello A., Grillo C., Jones T., Treu T., Bonamigo M., Suyu S.~H., 2018, MNRAS, 474, 3391
\bibitem[\protect\citeauthoryear{Agnello et al.}{2018c}]{agnello18c} Agnello A., et al., 2018, MNRAS, 479, 4345
\bibitem[\protect\citeauthoryear{Agnello \& Spiniello}{2018}]{agnellospiniello18} Agnello A., Spiniello C., 2018, arXiv, arXiv:1805.11103 
\bibitem[\protect\citeauthoryear{Aihara et al.}{2018}]{aihara18} Aihara H., et al., 2018, PASJ, 70, S8 
\bibitem[\protect\citeauthoryear{Arenou et al.}{2018}]{arenou18} Arenou F., et al., 2018, A\&A, 616, A17 
\bibitem[\protect\citeauthoryear{Astropy Collaboration et al.}{2013}]{astropy13} Astropy Collaboration, et al., 2013, A\&A, 558, A33 
\bibitem[\protect\citeauthoryear{Auger et al.}{2009}]{auger09} Auger M.~W., Treu T., Bolton A.~S., Gavazzi R., Koopmans L.~V.~E., Marshall P.~J., Bundy K., Moustakas L.~A., 2009, ApJ, 705, 1099 
\bibitem[\protect\citeauthoryear{Ben{\'{\i}}tez}{2000}]{benitez00} Ben{\'{\i}}tez N., 2000, ApJ, 536, 571
\bibitem[\protect\citeauthoryear{Berghea et al.}{2017}]{berghea17} Berghea C.~T., Nelson G.~J., Rusu C.~E., Keeton C.~R., Dudik R.~P., 2017, ApJ, 844, 90
\bibitem[\protect\citeauthoryear{Bertin 
\& Arnouts}{1996}]{bertin96} Bertin E., Arnouts S., 1996, A\&AS, 117, 393 
\bibitem[\protect\citeauthoryear{Berlind et al.}{2006}]{berlind06} Berlind A.~A., et al., 2006, ApJS, 167, 1
\bibitem[\protect\citeauthoryear{Bertin et al.}{2002}]{bertin02} Bertin E., Mellier Y., Radovich M., Missonnier G., Didelon P., Morin B., 2002, ASPC, 281, 228 
\bibitem[\protect\citeauthoryear{Bertin}{2006}]{bertin06} Bertin E., 2006, ASPC, 351, 112 
\bibitem[\protect\citeauthoryear{Bettoni et al.}{2019}]{bettoni19} Bettoni D., Falomo R., Scarpa R., Negrello M., Omizzolo A., Corradi R.~L.~M., Reverte D., Vulcani B., 2019, arXiv, arXiv:1902.10964 
\bibitem[\protect\citeauthoryear{Bonvin et al.}{2017}]{bonvin17} Bonvin V., et al., 2017, MNRAS, 465, 4914
\bibitem[\protect\citeauthoryear{Cardelli, Clayton, \& Mathis}{1989}]{cardelli89} Cardelli J.~A., Clayton G.~C., Mathis J.~S., 1989, ApJ, 345, 245 
\bibitem[\protect\citeauthoryear{Chambers et al.}{2016}]{chambers16} Chambers K.~C., et al., 2016, arXiv, arXiv:1612.05560 
\bibitem[\protect\citeauthoryear{Chan et al.}{2015}]{chan15} Chan J.~H.~H., Suyu S.~H., Chiueh T., More A., Marshall P.~J., Coupon J., Oguri M., Price P., 2015, ApJ, 807, 138
\bibitem[\protect\citeauthoryear{Chen et al.}{2016}]{chen16} Chen G.~C.-F., et al., 2016, MNRAS, 462, 3457
\bibitem[\protect\citeauthoryear{Claeskens \& Surdej}{2002}]{claeskens02} Claeskens J.-F., Surdej J., 2002, A\&ARv, 10, 263 
\bibitem[\protect\citeauthoryear{Colless et al.}{2001}]{colless01} Colless M., et al., 2001, MNRAS, 328, 1039 
\bibitem[\protect\citeauthoryear{Croom et al.}{2001}]{croom01} Croom S.~M., Smith R.~J., Boyle B.~J., Shanks T., Loaring N.~S., Miller L., Lewis I.~J., 2001, MNRAS, 322, L29 
\bibitem[\protect\citeauthoryear{D'Abrusco et al.}{2014}]{d'abrusco14} D'Abrusco R., Massaro F., Paggi A., Smith H.~A., Masetti N., Landoni M., Tosti G., 2014, ApJS, 215, 14 
\bibitem[\protect\citeauthoryear{da Cunha et al.}{2017}]{dacunha17} da Cunha E., et al., 2017, PASA, 34, e047 
\bibitem[\protect\citeauthoryear{Delchambre et al.}{2019}]{delchambre18} Delchambre L., et al., 2019, A\&A, 622, A165
\bibitem[\protect\citeauthoryear{de Vaucouleurs}{1948}]{devaucouleurs48} de Vaucouleurs G., 1948, AnAp, 11, 247
\bibitem[\protect\citeauthoryear{Faber \& Jackson}{1976}]{faber76} Faber S.~M., Jackson R.~E., 1976, ApJ, 204, 668 
\bibitem[\protect\citeauthoryear{Fabricius et al.}{2016}]{fabricius16} Fabricius C., et al., 2016, A\&A, 595, A3 
\bibitem[\protect\citeauthoryear{Falco et al.}{1999}]{falco99} Falco E.~E., et al., 1999, ApJ, 523, 617
\bibitem[\protect\citeauthoryear{Findlay et al.}{2018}]{findlay18} Findlay J.~R., et al., 2018, ApJS, 236, 44
\bibitem[\protect\citeauthoryear{Flaugher et al.}{2015}]{flaugher15} Flaugher B., et al., 2015, AJ, 150, 150 
\bibitem[\protect\citeauthoryear{Flesch}{2015}]{flesch15} Flesch E.~W., 2015, PASA, 32, e010 
 \bibitem[\protect\citeauthoryear{Flewelling et al.}{2016}]{flewelling16} Flewelling H.~A., et al., 2016, arXiv, arXiv:1612.05243
\bibitem[\protect\citeauthoryear{Foreman-Mackey et al.}{2013}]{foreman13} Foreman-Mackey D., Hogg D.~W., Lang D., Goodman J., 2013, PASP, 125, 306 
\bibitem[\protect\citeauthoryear{Gaia Collaboration et al.}{2016}]{gaia16} Gaia Collaboration, et al., 2016, A\&A, 595, A1
\bibitem[\protect\citeauthoryear{Gaia Collaboration et al.}{2018a}]{gaia18a} Gaia Collaboration, et al., 2018, A\&A, 616, A1 
\bibitem[\protect\citeauthoryear{Gaia Collaboration et al.}{2018b}]{gaia18b} Gaia Collaboration, et al., 2018, A\&A, 616, A14 
\bibitem[\protect\citeauthoryear{Healey et al.}{2008}]{healey08} Healey S.~E., et al., 2008, ApJS, 175, 97 
\bibitem[\protect\citeauthoryear{Hennawi et al.}{2006}]{hennawi06} Hennawi J.~F., et al., 2006, AJ, 131, 1 
\bibitem[\protect\citeauthoryear{Hennawi et al.}{2010}]{hennawi10} Hennawi J.~F., et al., 2010, ApJ, 719, 1672 
\bibitem[\protect\citeauthoryear{Henstock et al.}{1997}]{henstock97} Henstock D.~R., Browne I.~W.~A., Wilkinson P.~N., McMahon R.~G., 1997, MNRAS, 290, 380 
\bibitem[\protect\citeauthoryear{Hewett et al.}{1994}]{hewett94} Hewett P.~C., Irwin M.~J., Foltz C.~B., Harding M.~E., Corrigan R.~T., Webster R.~L., Dinshaw N., 1994, AJ, 108, 1534
\bibitem[\protect\citeauthoryear{Holder \& Schechter}{2003}]{holder03} Holder G.~P., Schechter P.~L., 2003, ApJ, 589, 688 
\bibitem[\protect\citeauthoryear{Hopkins et al.}{2006}]{hopkins06} Hopkins P.~F., Hernquist L., Cox T.~J., Di Matteo T., Robertson B., Springel V., 2006, ApJS, 163, 1 
\bibitem[\protect\citeauthoryear{Hopkins et al.}{2008}]{hopkins08} Hopkins P.~F., Hernquist L., Cox T.~J., Kere{\v s} D., 2008, ApJS, 175, 356 
\bibitem[\protect\citeauthoryear{Hunter}{2007}]{hunter07} Hunter, J.~D., et al. 2007, Computing in Science \& Engineering, 9, 3, 90-95
\bibitem[\protect\citeauthoryear{Huo et al.}{2013}]{huo13} Huo Z.-Y., et al., 2013, AJ, 145, 159
\bibitem[\protect\citeauthoryear{Inada et al.}{2003}]{inada03} Inada N., et al., 2003, Natur, 426, 810
\bibitem[\protect\citeauthoryear{Inada et al.}{2008}]{inada08} Inada N., et al., 2008, AJ, 135, 496
\bibitem[\protect\citeauthoryear{Inada et al.}{2009}]{inada09} Inada N., et al., 2009, AJ, 137, 4118
\bibitem[\protect\citeauthoryear{Inada et al.}{2010}]{inada10} Inada N., et al., 2010, AJ, 140, 403 
\bibitem[\protect\citeauthoryear{Inada et al.}{2012}]{inada12} Inada N., et al., 2012, AJ, 143, 119 
\bibitem[\protect\citeauthoryear{Inada et al.}{2014}]{inada14} Inada N., Oguri M., Rusu C.~E., Kayo I., Morokuma T., 2014, AJ, 147, 153
\bibitem[\protect\citeauthoryear{Jackson et al.}{2012}]{jackson12} Jackson N., Rampadarath H., Ofek E.~O., Oguri M., Shin M.-S., 2012, MNRAS, 419, 2014 
\bibitem[\protect\citeauthoryear{Jean \& Surdej}{1998}]{jean98} Jean C., Surdej J., 1998, A\&A, 339, 729
\bibitem[\protect\citeauthoryear{Jones et al.}{2009}]{jones09} Jones D.~H., et al., 2009, MNRAS, 399, 683 
\bibitem[\protect\citeauthoryear{Kayo et al.}{2010}]{kayo10} Kayo I., Inada N., Oguri M., Morokuma T., Hall P.~B., Kochanek C.~S., Schneider D.~P., 2010, AJ, 139, 1614 
\bibitem[\protect\citeauthoryear{Kayo \& Oguri}{2012}]{kayo12} Kayo I., Oguri M., 2012, MNRAS, 424, 1363 
\bibitem[\protect\citeauthoryear{Keeton, Kochanek, \& Falco}{1998}]{keeton98} Keeton C.~R., Kochanek C.~S., Falco E.~E., 1998, ApJ, 509, 561
\bibitem[\protect\citeauthoryear{Kleinman et al.}{2013}]{kleinman13} Kleinman S.~J., et al., 2013, ApJS, 204, 5 
\bibitem[\protect\citeauthoryear{Kneib, Cohen, \& Hjorth}{2000}]{kneib00} Kneib J.-P., Cohen J.~G., Hjorth J., 2000, ApJ, 544, L35 
\bibitem[\protect\citeauthoryear{Koposov, Belokurov, \& Torrealba}{2017}]{koposov17} Koposov S.~E., Belokurov V., Torrealba G., 2017, MNRAS, 470, 2702 
\bibitem[\protect\citeauthoryear{Kostrzewa-Rutkowska et al.}{2018}]{kostrzewa18} Kostrzewa-Rutkowska Z., et al., 2018, MNRAS
\bibitem[\protect\citeauthoryear{Lee}{2017}]{lee17} Lee C.-H., 2017, A\&A, 605, L8 
\bibitem[\protect\citeauthoryear{Lee}{2018}]{lee18} Lee C.-H., 2018, MNRAS, 475, 3086 
\bibitem[\protect\citeauthoryear{Lemon et al.}{2017}]{lemon17} Lemon C.~A., Auger M.~W., McMahon R.~G., Koposov S.~E., 2017, MNRAS, 472, 5023
\bibitem[Lemon et al.(2018)]{lemon18} Lemon, C.~A., Auger, M.~W., McMahon, R.~G., \& Ostrovski, F.\ 2018, \mnras, 479, 5060 
\bibitem[\protect\citeauthoryear{Lemon, Auger, \& McMahon}{2019}]{lemon19} Lemon C.~A., Auger M.~W., McMahon R.~G., 2019, MNRAS, 483, 4242
\bibitem[\protect\citeauthoryear{Lindegren et al.}{2016}]{lindegren16} Lindegren L., et al., 2016, A\&A, 595, A4 
\bibitem[\protect\citeauthoryear{Lucey et al.}{2018}]{lucey18} Lucey J.~R., Schechter P.~L., Smith R.~J., Anguita T., 2018, MNRAS, 476, 927 
\bibitem[\protect\citeauthoryear{Maddox et al.}{1990}]{maddox90} Maddox S.~J., Sutherland W.~J., Efstathiou G., Loveday J., 1990, MNRAS, 243, 692 
\bibitem[\protect\citeauthoryear{Magain et al.}{2005}]{magain05} Magain P., Letawe G., Courbin F., Jablonka P., Jahnke K., Meylan G., Wisotzki L., 2005, Natur, 437, 381
\bibitem[\protect\citeauthoryear{Magnier et al.}{2016a}]{magnier16a} Magnier E.~A., et al., 2016a, arXiv, arXiv:1612.05240 
\bibitem[\protect\citeauthoryear{Magnier et al.}{2016b}]{magnier16b} Magnier E.~A., et al., 2016b, arXiv, arXiv:1612.05242 
\bibitem[\protect\citeauthoryear{McMahon et al.}{2013}]{mcmahon13} McMahon R.~G., Banerji M., Gonzalez E., Koposov S.~E., Bejar V.~J., Lodieu N., Rebolo R., VHS Collaboration, 2013, Msngr, 154, 35
\bibitem[\protect\citeauthoryear{Moffat}{1969}]{moffat69} Moffat A.~F.~J., 1969, A\&A, 3, 455
\bibitem[\protect\citeauthoryear{More et al.}{2016}]{more16} More A., et al., 2016, MNRAS, 456, 1595
\bibitem[\protect\citeauthoryear{Nelder \& Mead}{1965}]{nelder65} Nelder, J.~A., Mead. R., 1965, The Computer Journal 7, 308-13
\bibitem[\protect\citeauthoryear{Ochner et al.}{2014}]{ochner14} Ochner P., et al., 2014, ATel, 6750
\bibitem[\protect\citeauthoryear{Ofek et al.}{2007}]{ofek07} Ofek E.~O., Oguri M., Jackson N., Inada N., Kayo I., 2007, MNRAS, 382, 412
\bibitem[\protect\citeauthoryear{Oguri et al.}{2005}]{oguri05} Oguri M., et al., 2005, ApJ, 622, 106 
\bibitem[\protect\citeauthoryear{Oguri et al.}{2006}]{oguri06} Oguri M., et al., 2006, AJ, 132, 999
\bibitem[\protect\citeauthoryear{Oguri et al.}{2008}]{oguri08} Oguri M., et al., 2008, AJ, 135, 520 
\bibitem[\protect\citeauthoryear{Oguri}{2010}]{oguri10a} Oguri M., 2010, PASJ, 62, 1017
\bibitem[\protect\citeauthoryear{Oguri \& Marshall}{2010}]{oguri10b} Oguri M., Marshall P.~J., 2010, MNRAS, 405, 2579 
\bibitem[\protect\citeauthoryear{Oguri et al.}{2012}]{oguri12} Oguri M., et al., 2012, AJ, 143, 120
\bibitem[\protect\citeauthoryear{Oliphant}{2007}]{oliphant07} Oliphant, T.~E., 2007, Computing in Science \& Engineering, 9, 10-20
\bibitem[Onaka \& al.(2008)]{ona08} Onaka P., Tonry J.~L., Isani S., Lee A., Uyeshiro R., Rae C., Robertson L., Ching G., Proc.\ 2008, \procspie, 7014, 12
\bibitem[\protect\citeauthoryear{Ostrovski et al.}{2017}]{ostrovski17} Ostrovski F., et al., 2017, MNRAS, 465, 4325
\bibitem[\protect\citeauthoryear{Ostrovski et al.}{2018}]{ostrovski18} Ostrovski F., et al., 2018, MNRAS, 473, L116 
\bibitem[\protect\citeauthoryear{Peng et al.}{1999}]{peng99} Peng C.~Y., et al., 1999, ApJ, 524, 572 
\bibitem[\protect\citeauthoryear{Rubin et al.}{2018}]{rubin18} Rubin K.~H.~R., et al., 2018, ApJ, 859, 146
\bibitem[\protect\citeauthoryear{Rusu et al.}{2013}]{rusu12} Rusu C.~E., Oguri M., Iye M., Inada N., Kayo I., Shin M.-S., Sluse D., Strauss M.~A., 2013, ApJ, 765, 139 
\bibitem[\protect\citeauthoryear{Rusu et al.}{2016}]{rusu16} Rusu C.~E., et al., 2016, MNRAS, 458, 2
 \bibitem[\protect\citeauthoryear{Rusu et al.}{2017}]{rusu17} Rusu C.~E., et al., 2017, MNRAS, 467, 4220 
 \bibitem[\protect\citeauthoryear{Rusu \& Lemon}{2018}]{rusu18} Rusu C.~E., Lemon C.~A., 2018, RNAAS, 2, 187 
 \bibitem[\protect\citeauthoryear{Sergeyev et al.}{2016}]{sergeyev16} Sergeyev A.~V., Zheleznyak A.~P., Shalyapin V.~N., Goicoechea L.~J., 2016, MNRAS, 456, 1948
 \bibitem[\protect\citeauthoryear{Schechter et al.}{2014}]{schechter14} Schechter P.~L., Pooley D., Blackburne J.~A., Wambsganss J., 2014, ApJ, 793, 96 
\bibitem[\protect\citeauthoryear{Schlafly \& Finkbeiner}{2011}]{schlafly11} Schlafly E.~F., Finkbeiner D.~P., 2011, ApJ, 737, 103 
\bibitem[Secrest et al.(2015)]{secrest15} Secrest, N.~J., Dudik, R.~P., Dorland, B.~N., et al.\ 2015, \apjs, 221, 12 
\bibitem[\protect\citeauthoryear{Shanks et al.}{2015}]{shanks15} Shanks T., et al., 2015, MNRAS, 451, 4238
\bibitem[\protect\citeauthoryear{Sheinis et al.}{2002}]{ESI} Sheinis A.~I., Bolte M., Epps H.~W., Kibrick R.~I., Miller J.~S., Radovan M.~V., Bigelow B.~C., Sutin B.~M., 2002, PASP, 114, 851 
\bibitem[Sheth et al.(2003)]{sheth03} Sheth, R.~K., Bernardi, M., Schechter, P.~L., et al.\ 2003, \apj, 594, 225
\bibitem[\protect\citeauthoryear{Skrutskie et al.}{1997}]{skrutskie97} Skrutskie M.~F., et al., 1997, ASSL, 210, 25 
\bibitem[\protect\citeauthoryear{Sluse et al.}{2012}]{sluse12} Sluse D., Chantry V., Magain P., Courbin F., Meylan G., 2012, A\&A, 538, A99 
\bibitem[\protect\citeauthoryear{Sluse et al.}{2017}]{sluse17} Sluse D., et al., 2017, MNRAS, 470, 4838
\bibitem[\protect\citeauthoryear{Spiniello et al.}{2018}]{spiniello18} Spiniello C., et al., 2018, MNRAS, 480, 1163
\bibitem[\protect\citeauthoryear{Surdej et al.}{1987}]{surdej87} Surdej J., et al., 1987, Natur, 329, 695 
\bibitem[\protect\citeauthoryear{Szabo et al.}{2011}]{szabo11} Szabo T., Pierpaoli E., Dong F., Pipino A., Gunn J., 2011, ApJ, 736, 21 
\bibitem[\protect\citeauthoryear{Taylor}{2005}]{taylor05} Taylor M.~B., 2005, ASPC, 347, 29 
\bibitem[Tonry \& Onaka(2008)]{ton09} Tonry J., Onaka P.\ 2009, in Ryan S., ed., 
Proceedings of the Advanced Maui Optical and Space Surveillance Technologies Conference.
The Maui Economic Development Board, Kihei, HI, p. E40
\bibitem[\protect\citeauthoryear{Treu \& Marshall}{2016}]{treu16} Treu T., Marshall P.~J., 2016, A\&ARv, 24, 11  
\bibitem[\protect\citeauthoryear{Treu et al.}{2018}]{treu18} Treu T., et al., 2018, MNRAS, 481, 1041 
\bibitem[\protect\citeauthoryear{Weymann et al.}{1980}]{weymann80} Weymann R.~J., Latham D., Angel J.~R.~P., Green R.~F., Liebert J.~W., Turnshek D.~A., Turnshek D.~E., Tyson J.~A., 1980, Natur, 285, 641
\bibitem[\protect\citeauthoryear{Williams, Agnello, \& Treu}{2017}]{williams17} Williams P., Agnello A., Treu T., 2017, MNRAS, 466, 3088 
\bibitem[\protect\citeauthoryear{Wilson et al.}{2016}]{wilson16} Wilson M.~L., Zabludoff A.~I., Ammons S.~M., Momcheva I.~G., Williams K.~A., Keeton C.~R., 2016, ApJ, 833, 194
\bibitem[\protect\citeauthoryear{Wisotzki et al.}{1993}]{wisotzki93} Wisotzki L., Koehler T., Kayser R., Reimers D., 1993, A\&A, 278, L15 
\bibitem[Wisotzki et al.(2002)]{wisotzki02} Wisotzki, L., Schechter, P.~L., Bradt, H.~V., Heinm{\"u}ller, J., \& Reimers, D.\ 2002, \aap, 395, 17 
\bibitem[\protect\citeauthoryear{Wong et al.}{2017}]{wong17} Wong K.~C., et al., 2017, MNRAS, 465, 4895 
\bibitem[\protect\citeauthoryear{Wright et al.}{2010}]{wright10} Wright E.~L., et al., 2010, AJ, 140, 1868-1881 
\bibitem[\protect\citeauthoryear{Yonehara, Hirashita, \& Richter}{2008}]{yonehara08} Yonehara A., Hirashita H., Richter P., 2008, A\&A, 478, 95
\bibitem[\protect\citeauthoryear{York et al.}{2000}]{york00} York D.~G., et al., 2000, AJ, 120, 1579 

\end{thebibliography}


\appendix

\section{Expected sample purity from the relative density of $Gaia$ and AGN sources}
\label{sec:appendix1} 

We start with the complete $Gaia$ DR2 source catalogue\footnote{\url{https://www.astro.rug.nl/~gaia/}}, where we apply the same automatic selection cuts we used for our sample of candidates. We use a bright magnitude limit of $G\geq17.5$, slightly brighter than our candidate source companions, and a faint one of $G\leq20$. While 29 of the 91 candidate source companions are fainter than $G=20$, we apply this cut because $Gaia$ DR2 is still complete at this limit \citep{arenou18}, outside the crowded regions excluded by our galactic latitude cut; the completeness is expected to drop towards the limiting magnitude of $G\sim21$. This reduces the $Gaia$ catalogue to $\sim148$ million sources. We also apply a color cut of $0\leq Rp-Bp\leq 1.5$, from Figure~\ref{fig:colormag}, as well as the same $Gaia$-based astrometric quantity cuts from Section~\ref{sec:searchgaia}. As we did for our candidates, we keep the objects without Gaia astrometric quantities or color. This results in $\sim22$ million remaining sources. 

The \citet{secrest15} AGN catalogue is less complete, but not significantly so, with a limiting magnitude of 20 in $g-$band, or about the same in $Gaia$ $G-$band\footnote{From our 312 candidates, $G-g$ has a distribution with a median of -0.21 and a standard deviation of 0.34}. We use the Milliquas catalogue \citep{flesch15}, which is slightly more complete at this magnitude limit (see Figure 3 in \citet{lemon19}), and includes high-confidence quasars detected in X-ray and radio, in addition to $WISE$. We cross-matched the $Gaia$ catalogue after performing the cuts described above with the Milliquas catalogue, resulting in $\sim520000$ matches. 

Finally, the $Gaia$ resolution is much higher than the one of $WISE$, with PSF FWHM $\sim6\arcsec$ \citep{wright10}. We use \texttt{Topcat} \citep{taylor05} to identify all objects from the $Gaia$-based catalogue we produced above with relative separation less than the $WISE$ PSF FWHM, and count each of these clusters as one. This reduces the $Gaia$-based sample to $\sim14$ million sources, resulting in an AGN fraction of $\sim4\%$. 

\section{Previously confirmed or ruled out candidates}
\label{sec:appendix}

\begin{table*}
  \scriptsize
 \centering
 \begin{minipage}{\linewidth}
   \caption{Previously confirmed candidates, and those ruled out by $Gaia$ data or existing spectroscopy}
  \begin{tabular}{@{}lrrclccl@{}}
  \hline 
Name [PS1~J...] & $\alpha$ & $\delta$ & \#Comp & $i$ & Sep. [\arcsec] & Rank & Notes \\ 
 \hline
013459+243049 & 23.745589 & $24.513635$     & 2  &  19.40   & 3.7              & C & $G=$19.60, 20.17; p-l; two SDSS QSOs at z=2.093 and z=2.104\\
014710+463043 & 26.792452 & $46.512081$     & 2  &  15.57   & 3.2              & A & $G=$15.89, 16.18; 16.74, 18.26; similar color p-l sources; PSJ0147+4630 \\
& &  &  &  & & & \citep[quad;][]{berghea17}\\
024526-055700 & 41.356685 & $-5.950128$     & 2  &  18.67   & 1.7              & C & $G=$19.73, 19.25; similar color p-l + red inner component;  \\
& &  &  &  & & & DESJ0245-0556 \citep[double;][]{agnello18b}\\
025934-233802 & 44.889982 & $-23.633792$    & 2  &  19.21    & 2.7               & B & $G=$20.34, 19.37; p-l + red inner component; PSJ0259-2338 \\
& &  &  &  & & & \citep[double;][]{lemon18}\\
094235+231030 & 145.645825& $23.175133$     & 2  &  18.87   & 2.4             & C & $G=$19.10, 19.92; similar color p-l; z=1.83 QSO pair \citep{findlay18}\\
110633-182124 & 166.639282& $-18.356688$    & 2  &  16.95    & 3.1              & C & $G=$17.07, 18.20; similar color p-l; HE1104-1805 \\
& &  &  &  & & & \citep[double;][]{wisotzki93}\\
110932+531636 & 167.384487& $53.276552$     & 2  &  18.71   & 3.2              & C & $G=$18.97, 19.68; similar color p-l; includes SDSS z=0.982 QSO; \\
& &  &  &  & & & SQLS QSO pair\\
113441-210323 & 173.668953& $-21.056307$    & 3  &  16.81   & 3.7              & A & $G=$17.17, 17.19, 18.94, 17.27; four p-l sources; 2M1134-2103 \\
& &  &  &  & & & \citep[quad;][]{lucey18}\\
120451+442836 & 181.210712& $44.47659$      & 2  &  18.84   & 3.0              & B & $G=$18.81, 19.65; similar color p-l; SQLS QSO at z=1.84 and z=1.14\\
120630+433219 & 181.623684& $43.538734$     & 2  &  18.52   & 3.0               & B & $G=$18.87, 18.84; similar color p-l + red component; SDSSJ1206+4332\\
& &  &  &  & & & \citep[double;][]{oguri05} \\
120659-254331 & 181.744763& $-25.725376$     & 2  &  19.41     & 2.1              & B & 19.97, 20.40; similar color p-l; both have negligible AEN, pm and p; double,\\
& &  &  &  & & & discovered and confirmed independently by C. Lemon, private communication \\
124614+503049 & 191.556942& $50.513634$     & 2  &  19.22   & 2.4              & C & $G=$19.31, 19.50; similar color p-l; SDSS quasar pair z=2.73, 2.11\\
132100+164403 & 200.246658& $16.734072$     & 3  &  18.51  & 8.8              & B & $G=$18.66, 19.46; similar color p-l + red component; SDSSJ1320+1644\\
& &  &  &  & & &  \citep[double or binary quasar;][]{rusu12} \\
133713+601208 & 204.304581& $60.202141$     & 2  &  18.55   & 3.1              & C & $G=$18.68, 19.63; similar color p-l;  z=1.721, 1.726 QSO pair \\
& &  &  &  & & & \citep{hennawi06}\\
140515+095930 & 211.314397& $9.991796$      & 2  &  18.96   & 1.9               & B & $G=$19.38, 20.32; p-l + extended red; ULAS J1405+0959 \\
& &  &  &  & & & \citep[double][]{jackson12}\\
141818-161008 & 214.573673& $-16.168771$    & 2  &  18.53   & 2.4              & C & $G=$18.46, 19.33; similar color p-l; NIQ z=1.13 \\
& &  &  &  & & & \citep{lemon19}\\
143323+600715 & 218.345158& $60.120864$     & 5  &  19.49   & 3.7           & A & $G=$19.87, 19.99, 20.26; p-l; SDSSJ1433+6007 \\
& &  &  &  & & & \citep[quad;][]{agnello18a}\\
143351+145007 & 218.462505& $14.835308$     & 2  &  18.90   & 3.3              & C & $G=$18.99, 19.35; similar color p-l; z=1.51 QSO pair \citep{findlay18}\\
151539+151135 & 228.910562& $15.193168$     & 2  &  17.94   & 2.0              & C & $G=$18.03, 18.42; similar color p-l; SDSSJ1515+1511 \\
& &  &  &  & & & \citep[double;][]{inada14}\\
153725-301017 & 234.355599& $-30.171336$    & 4  &  19.12   & 3.1               & A & $G=$20.32, 20.22, 20.44; four p-l + inner red component; \\
& &  &  &  & & & \citep[quad;][]{delchambre18,lemon19} \\
160600-233322 & 241.500981& $-23.556046$    & 3  &  17.96   & 2.9              & C & $G=$18.85, 18.97, 19.33, 19.61; p-l + red component; PSJ1606-2333 \\
& &  &  &  & & & \citep[quad;][]{lemon18}\\
172145+884222 & 260.43637 & $88.706169$     & 2  &  17.33   & 2.3             & B & $G=$18.18, 18.33; similar color; PSJ1721+8842 \citep[quad;][]{lemon18}\\
203238-235822 & 308.157206& $-23.972856$    & 2  &  18.75   & 2.0              & C & $G=$19.12, 19.26; similar color p-l; z=1.64 NIQ\\
& &  &  &  & & &  \citep{lemon18}\\
215316+273235 & 328.31765 & $27.543058$     & 2  &  18.69   & 3.6              & C & $G=$18.72, 19.69; similar color p-ls; quasar pair \citep{sergeyev16}\\
221208+314417 & 333.033412& $31.73809$      & 2  &  19.27   & 2.6              & C & $G=$19.28, 19.97; two similar color p-l + red component; \\
& &  &  &  & & & \citep[double;][]{lemon19}\\
\hline
000823+031342 & 2.094362  & $3.228219$      & 2  &  17.70   & 3.2              & C & p-l; PB 5757 (star), large pm\\
001313-152007 & 3.302628  & $-15.335383$    & 2  &  16.82   & 1.9              & C & similar color p-l; companion has large pm\\
002605+401519 & 6.522825  & $40.255255$     & 2  &  17.85   & 2.3              & C & p-l; companion has large pm\\
002719+300336 & 6.827338  & $30.059894$     & 2  &  17.90   & 3.0              & C & similar color p-l; companion has large pm\\
004346+282715 & 10.942056 & $28.454297$     & 2  &  17.58   & 3.3              & C & point sources; bright component has large pm and p\\
004446+472400 & 11.192613 & $47.399741$     & 2  &  18.13   & 3.3              & C & similar color p-l; companion has large pm\\
005801-231711 & 14.502676 & $-23.286411$    & 2  &  17.82   & 3.1              & C & similar color p-l; one component has large AEN\\
011305+454905 & 18.269259 & $45.818058$     & 2  &  17.60   & 1.7              & C & similar color p-l; companion has large pm\\
011639+405252 & 19.163546 & $40.881125$     & 2  &  18.72   & 1.3             & C & similar color p-l; includes SDSS z=1.86 QSO; one component has large p\\
015109+315521 & 27.786404 & $31.922389$     & 2  &  18.11   & 3.1             & C & includes galaxy \citep{ochner14}; companion has large pm\\
020122+212637 & 30.340775 & $21.443685$     & 2  &  17.45   & 3.6              & C & similar color p-l; SQLS candidate; companion has large pm\\
020649+803347 & 31.703677 & $80.563065$     & 2  &  17.36   & 2.2              & C & similar color p-l; companion has large pm and p\\
020722+374720 & 31.843188 & $37.788868$     & 2  &  16.60   & 2.2             & C & similar color p-l + extended; companion has large pm and p\\
024414-073747 & 41.059791 & $-7.629853$     & 2  &  19.75   & 1.4              & C & p-l; includes z=0.319 galaxy \citep{szabo11}; companion has no \\
& &  &  &  & & &Gaia pm and p\\
024722-172547 & 41.843352 & $-17.429683$    & 2  &  18.67   & 1.6              & C & similar color p-l; companion has large pm\\
025339+070440 & 43.414166 & $7.077896$      & 2  &  18.17   & 2.8             & C & similar color p-l; companion has large pm\\
025644+394153 & 44.183373 & $39.697932$     & 2  &  19.28    & 2.4              & C & similar color p-l; companion has large pm\\
034955-071723 & 57.479613 & $-7.289607$     & 2  &  16.12   & 2.5               & C & p-l; companion has large pm and p\\
035119-182302 & 57.829409 & $-18.383904$    & 2  &  16.92   & 1.8              & C & p-l; companion has large pm and p\\
041304+155206 & 63.266175 & $15.868444$     & 2  &  17.60   & 3.5               & C & similar color p-l; companion has large pm\\
043324-111537 & 68.348462 & $-11.260161$    & 2  &  18.31   & 2.7              & C & similar color p-l; companion has large pm\\
045230-295335 & 73.125436 & $-29.893138$    & 2  &  15.36   & 2.0              & B & similar color p-l + extended? star+interacting galaxy+QSO \\
& &  &  &  & & & HE0450-2958 \citep{magain05}\\
051139-035102 & 77.911071 & $-3.850553$     & 2  &  19.02   & 2.9              & C & similar color p-l; QSO+other \citep{lemon18}; both negligible \\
& &  &  &  & & & AEN, pm and p\\
052131+730136 & 80.380313 & $73.026614$     & 2  &  17.55     & 3.0              & C & p-l; companion has large pm and p\\
052419-065727 & 81.077162 & $-6.957592$     & 2  &  16.98   & 1.8             & C & p-l; companion has large pm\\
052833+042744 & 82.136604 & $4.462234$      & 2  &  16.72    & 1.9              & C & p-l; companion has large pm\\
053733+815634 & 84.386019 & $81.942802$     & 2  &  18.72   & 3.2              & C & similar color p-l; companion has large pm\\
054335-152624 & 85.894214 & $-15.439864$    & 2  &  17.63     & 2.8             & C & p-l; companion has large pm\\
061050-201839 & 92.710135 & $-20.310915$    & 2  &  18.37   & 2.8              & C & p-l; companion has large pm\\
061911-295857 & 94.796622 & $-29.982405$    & 3  &  18.70   & 2.3               & C & different color p-l; outer component has large AEN and pm; \\
& &  &  &  & & & included in the \citet{delchambre18} Gaia clusters catalogue\\
062529-285546 & 96.371981 & $-28.929452$    & 2  &  18.25   & 3.0             & C & p-l; companion has large pm\\
063724+434603 & 99.34909  & $43.767531$     & 3  &  16.9   & 5.3                   & C & similar color p-l (companion has large pm and p) + red inner component\\
065513+850519 & 103.804667& $85.088737$     & 3  &  17.50   & 4.5              & B & similar color p-l (companion has large pm) + red inner component \\
& &  &  &  & & &(large AEN); included in the \citet{delchambre18} Gaia clusters catalogue\\
\hline
\end{tabular}
\\ 
\end{minipage}
\end{table*}
\normalsize

\begin{table*}
  \scriptsize
 \centering
 \begin{minipage}{\linewidth}
 \contcaption{}
  \begin{tabular}{@{}lrrclccl@{}}
  \hline 
Name [PS1~J...] & $\alpha$ & $\delta$ & \#Comp & $i$ & Sep. [\arcsec] & Rank & notes \\ 
 \hline
072846+420701 & 112.190784& $42.116988$     & 2  &  16.79   & 3.9              & C & similar color p-l; includes SDSS z=1.120 QSO; SQLS candidate; \\
& &  &  &  & & & companion has large pm and p\\
072850+570125 & 112.206878& $57.02358$      & 2  &  16.03   & 3.5              & C & p-l; includes z=0.426 Seyfert 1 \citep{henstock97}; \\
& &  &  &  & & & companion has large pm and p\\
074242+651038 & 115.673761& $65.177097$     & 2  &  15.17   & 2.0              & C & consistent with single extended source; Mrk 78 (Seyfert 2); no Gaia data \\
074555+181818 & 116.478082& $18.304882$     & 2  &  17.80   & 2.5              & C & p-l; includes SDSS z=1.060 QSO; SQLS candidate; \\
& &  &  &  & & & companion has large pm\\
080938+275648 & 122.407538& $27.946714$     & 2  &  17.09   & 3.1             & C & similar color p-l; includes SDSS z=0.406 QSO; \\
& &  &  &  & & & companion has large pm and p\\
081130+255541 & 122.876253& $25.927955$     & 2  &  18.96   & 2.7              & C & p-l; companion has large pm\\
082218+665957 & 125.574156& $66.999183$     & 2  &  18.61   & 3.0               & C & similar color p-l; QSO+other \citep{lemon18}; \\
& &  &  &  & & &both negligible AEN, pm and p \\
082353-085114 & 125.970487& $-8.853931$     & 2  &  16.88   & 2.2              & C & similar color p-l; companion has large pm and p\\
082442+592409 & 126.176996& $59.402484$     & 2  &  17.77   & 3.1              & C & similar color p-l; companion has large pm\\
083229+563235 & 128.119012& $56.542997$     & 3  &  18.76    & 3.0              & C & similar color p-l (companion has large pm) + inner red component; \\
& &  &  &  & & & includes SDSS z=0.683 QSO; SQLS  QSO+star\\
084441+334909 & 131.16938 & $33.819226$     & 2  &  18.28   & 2.9              & C & similar color p-l; includes SDSS z=1.425 QSO; SQLS candidate; \\
& &  &  &  & & & companion has large pm \\
084513+543422 & 131.302961& $54.57264$      & 2  &  18.51   & 1.4              & C & similar color p-l; includes SDSS z=1.290 QSO; SQLS QSO+star (large pm)\\
085055-052735 & 132.72728 & $-5.459747$     & 2  &  19.11   & 2.0              & C & similar color p-l; companion has large pm\\
085838-152907 & 134.656619& $-15.485172$    & 2  &  17.27   & 2.5              & C & similar color p-l; companion has large pm\\
090852+304332 & 137.215844& $30.725594$     & 2  &  18.51   & 2.1                   & C & p-l; includes SDSS z=0.399 Seyfert 1; companion has large pm\\
091453-265223 & 138.722422& $-26.873106$    & 2  &  17.80   & 2.4              & C & similar color p-l; companion has large pm\\
091746-160623 & 139.443706& $-16.106479$    & 4  &  18.41   & 2.3              & C & similar color p-l; companion has large pm;\\
& &  &  &  & & &  included in the \citet{delchambre18} Gaia clusters catalogue\\
092016-063144 & 140.064718& $-6.529$        & 2  &  18.16   & 2.7              & C & similar color p-l; companion has large pm\\
092438-012845 & 141.157105& $-1.479089$     & 2  &  17.99   & 3.0              & C & similar color p-l; includes SDSS z=2.446 QSO; companion has large pm\\
092718+211357 & 141.826656& $21.232549$     & 2  &  17.47    & 2.3              & C & p-l; includes SDSS z=1.851 QSO; SQLS candidate, no lensing object; \\
& &  &  &  & & & companion has large pm \\
094115+305810 & 145.314113& $30.969479$     & 2  &  19.29   & 2.3              & C & similar color; SQLS z=1.193 QSO+blue galaxy\\
094437-263355 & 146.154045& $-26.565394$    & 2  &  16.77   & 2.3              & C & similar color; includes Seyfert 1 galaxy at z=0.142 \citep{jones09}\\
& &  &  &  & & & companion has large pm \\
094903+280022 & 147.264552& $28.006127$     & 2  &  18.79   & 1.2              & C & similar color p-l; SQLS QSO+star\\
100450+773753 & 151.208619& $77.63132$      & 2  &  19.05   & 1.9             & C & similar color p-l; companion has large pm\\
102803-153028 & 157.011143& $-15.507813$    & 3  &  19.60   & 4.5              & B & p-l + red inner component (large pm)\\
102813+171902 & 157.054777& $17.317297$     & 2  &  18.53   & 1.9              & C & p-l; only one component has Gaia p and pm, large pm\\
104704-241459 & 161.765852& $-24.249719$    & 2  &  16.95   & 2.8             & B & similar color p-l (companion has large pm) + red inner component\\
105852-275715 & 164.715138& $-27.954048$    & 2  &  18.01   & 2.4              & C & similar color p-l; companion has large pm\\
111524-042218 & 168.848654& $-4.371723$     & 2  &  18.60   & 2.7              & C & similar color p-l; includes galaxy at z=0.209 \citep{colless01}; \\
& &  &  &  & & & no Gaia p and pm, large AEN for companion \\
113431+111918 & 173.628607& $11.321701$     & 2  &  18.49   & 1.6              & C & similar color p-l; large companion pm; z=1.62 QSO+star \\
& &  &  &  & & & Ostrovski et al, in prep.\\
114214-075619 & 175.556357& $-7.93867$      & 2  &  18.93   & 3.4               & C & similar color p-l; companion has large pm\\
115443-224432 & 178.680182& $-22.742147$    & 2  &  17.75   & 2.4              & B & similar color p-l; companion has large pm\\
115541+131105 & 178.919792& $13.184774$     & 2  &  17.52    & 2.4              & B & similar color p-l; companion has large pm\\
115957+644406 & 179.987136& $64.735049$     & 2  &  18.77   & 3.1              & C & similar color p-l; includes SDSS z=1.61 QSO; companion has large pm\\
123441+341000 & 188.672008& $34.166556$     & 2  &  18.29   & 2.2             & C & similar color p-l; includes SDSS z=1.429 QSO, SQLS candidate; \\
& &  &  &  & & & companion has large pm\\
123559-023503 & 188.993809& $-2.58423$      & 2  &  17.78   & 3.0             & C & similar color p-l; SDSS z=2.062 QSO+star, SQLS candidate\\
130738+640252 & 196.907012& $64.047899$     & 2  &  18.17   & 3.5                & C & similar color p-l; companion has large pm\\
131425+181232 & 198.605024& $18.208753$     & 2  &  19.46   & 2.5              & C & p-l; companion has large pm\\
132223+512017 & 200.595155& $51.338029$     & 2  &  18.29   & 2.7              & C & similar color p-l; includes SDSS z=1.772 QSO; SQLS candidate; \\
& &  &  &  & & & companion has large pm\\
132405+282334 & 201.022027& $28.392698$     & 2  &  18.54   & 2.1              & C & similar color p-l; includes SDSS z=0.904 QSO; SQLS candidate, \\
& &  &  &  & & &  no lensing object; companion has large pm \\
132853+261501 & 202.222599& $26.250248$     & 2  &  18.91   & 2.6             & C & similar color p-l; SQLS candidate; SDSS z=1.522 QSO + star; \\
& &  &  &  & & & companion has large pm\\
132916+414554 & 202.31656 & $41.765054$     & 2  &  17.59   & 2.8             & C & similar color p-l; companion has large pm\\
133543-294239 & 203.927943& $-29.710967$    & 2  &  18.50   & 2.4              & C & p-l; companion has large pm\\
134222-261001 & 205.593589& $-26.166945$    & 2  &  18.30   & 2.9              & C & similar color p-l; companion has large pm\\
134539-262819 & 206.411024& $-26.471915$    & 2  &  17.82   & 2.6              & C & similar color p-l; companion has large pm\\
134626+045245 & 206.609217& $4.879294$      & 2  &  18.73   & 2.7              & C & similar color p-l; companion has large pm\\
134941+011054 & 207.420114& $1.181594$      & 2  &  16.55   & 2.2              & C & similar color p-l; includes SDSS star\\
140610-250809 & 211.540001& $-25.135907$    & 2  &  17.88   & 2.9              & B & similar color p-l; companion has large pm\\
141349+475113 & 213.452222& $47.853718$     & 2  &  18.55   & 3.0              & C & similar color p-l; includes SDSS z=2.175 QSO; SQLS candidate, \\
& &  &  &  & & &  no lensing object; companion has large pm \\
141432-052951 & 213.631386& $-5.49754$      & 2  &  19.17   & 2.2             & C & similar color p-l; companion has large pm\\
142040+122507 & 215.16569 & $12.418669$     & 2  &  18.31   & 3.1              & C & similar color p-l; includes SDSS z=2.252 QSO; companion has large pm\\
142402+710911 & 216.008966& $71.152985$     & 2  &  18.70   & 2.9              & C & similar color p-l; companion has large pm\\
142609-210327 & 216.538323& $-21.057381$    & 2  &  19.44   & 2.8                & C & similar color p-l; companion has large pm\\
143153-094341 & 217.972653& $-9.727974$     & 3  &  18.56   & 5.8              & B & p-l (companion has large pm) + red inner component\\
143154+530033 & 217.973863& $53.009266$     & 3  &  18.09   & 4.3               & C & p-l; includes SDSS z=1.389 QSO + star; third component has large pm\\
& &  &  &  & & & included in the \citet{delchambre18} Gaia clusters catalogue\\
143245-273713 & 218.188947& $-27.620192$    & 2  &  17.78   & 3.0              & C & similar color p-l; companion has large pm\\
144145+023743 & 220.437914& $2.628697$      & 2  &  19.13   & 1.1              & C & similar color p-l; SQLS z=1.160 QSO+star\\
144245+041619 & 220.689582& $4.271996$      & 2  &  19.27   & 3.0              & C & p-l; includes SDSS z=2.012 QSO; SQLS candidate; \\
& &  &  &  & & & companion has large pm\\
144303+260329 & 220.763978& $26.058137$     & 2  &  17.98   & 3.5              & C & similar color p-l; includes SDSS z=0.257 Seyfert 1;  \\
& &  &  &  & & & companion has large pm\\
\hline
\end{tabular}
\\ 
\end{minipage}
\end{table*}
\normalsize

\begin{table*}
  \scriptsize
 \centering
 \begin{minipage}{\linewidth}
 \contcaption{}
  \begin{tabular}{@{}lrrclccl@{}}
  \hline 
Name [PS1~J...] & $\alpha$ & $\delta$ & \#Comp & $i$ & Sep. [\arcsec] & Rank & notes \\ 
 \hline
 145115+052936 & 222.813312& $5.493197$      & 2  &  16.20   & 2.3             & C & similar color p-l; includes SDSS z=2.052 QSO; SQLS candidate; \\
& &  &  &  & & & companion has large pm\\
 145232-052947 & 223.134353& $-5.496432$     & 2  &  18.04   & 2.9              & C & p-l + red inner component; companion has large pm\\
145647-091751 & 224.197573& $-9.297562$     & 2  &  17.99   & 3.0              & C & similar color p-l; companion has large p and pm\\
150925+113851 & 227.35556 & $11.647604$     & 2  &  19.37   & 2.5               & C & similar color p-l; includes SDSS star; only one component has Gaia p \\
& &  &  &  & & & and pm, negligible values\\
151044-074043 & 227.684808& $-7.678621$     & 2  &  18.27  & 2.5              & C & similar color p-l; companion has large pm\\
151205+182706 & 228.018788& $18.451666$     & 2  &  17.76   & 2.5              & C & similar color p-l; companion has large pm\\
151237+553901 & 228.15381 & $55.650295$     & 2  &  19.01   & 1.9             & C & similar color p-l; includes SDSS z=1.363 QSO; SQLS QSO+star\\
151527-203609 & 228.862483& $-20.602366$    & 2  &  17.84   & 2.9              & C & p-l; companion has large pm\\
151832+343325 & 229.632917& $34.557016$     & 2  &  18.87   & 3.0              & C & similar color p-l; includes SDSS z=1.672 QSO; SQLS candidate; \\
& &  &  &  & & & companion has large pm\\
151858-022443 & 229.74139 & $-2.411924$     & 3  &  16.9   & 3.3              & B & p-l; inner component has large pm, the other two have negligible AEN, \\
& &  &  &  & & & p and pm\\
151918+094205 & 229.826754& $9.701277$      & 2  &  18.11   & 3.7              & C & similar color p-l; companion has large pm\\
152005+195038 & 230.021454& $19.843884$     & 3  &  18.73   & 1.4              & C & similar color; a single companion has Gaia data (only AEN, large)\\
152050+263741 & 230.209019& $26.627994$     & 2  &  18.93   & 2.1              & C & p-l; includes SDSS z=1.365 QSO; SQLS candidate candidate; \\
& &  &  &  & & & companion has large p and pm\\
152444+054628 & 231.182118& $5.77438$       & 2  &  17.68   & 3.7               & C & similar color p-l; includes SDSS z=1.445 QSO; SQLS candidate; \\
& &  &  &  & & & companion has large pm\\
153223-291257 & 233.094882& $-29.215933$    & 2  &  19.05   & 1.8              & C & similar color p-l; companion has large pm\\
153311-001509 & 233.296204& $-0.252541$     & 3  &  19.49   & 4.0              & C & similar colors; only one, outer component has Gaia data, large pm, the others \\
& &  &  &  & & & are galaxies\\
153510+082347 & 233.79015 & $8.396438$      & 2  &  18.77   & 2.5              & C & p-l; third component? includes SDSS z=1.953 QSO; SQLS candidate; \\
& &  &  &  & & & companion has large pm\\
154226-023456 & 235.610402& $-2.582135$     & 2  &  18.74   & 3.1              & C & similar color p-l; companion has large pm\\
154726-153237 & 236.857051& $-15.543614$    & 2  &  17.96    & 2.7              & C & similar color p-l; companion has large pm\\
160138+172852 & 240.407737& $17.48102$      & 3  &  17.80   & 5.6             & C & similar color p-l (one component has large pm) + red inner components\\
& &  &  &  & & & (large pm); includes SDSS z=2.239 QSO\\
160927+175431 & 242.361547& $17.90869$      & 2  &  17.67   & 2.6              & C & similar color p-l; includes SDSS z=1.993 QSO; SQLS candidate; \\
& &  &  &  & & & companion has large pm\\
161008+234837 & 242.533565& $23.810394$     & 2  &  19.14   & 3.0                & C & similar color p-l; companion has large pm\\
161657-170647 & 244.235836& $-17.113065$    & 2  &  18.12   & 2.6              & C & similar color p-l; companion has large pm\\
161722-230546 & 244.340087& $-23.096165$    & 2  &  18.80  & 1.8              & C & similar color p-l; QSO+star \citep{lemon18}\\
161841+301311 & 244.669946& $30.2196$       & 2  &  17.61   & 2.5              & B & similar color p-l; includes SDSS z=1.403 QSO; companion has large pm\\
161931+162123 & 244.878602& $16.356363$     & 2  &  19.08   & 2.6              & C & similar color p-l; includes SDSS z=2.455 QSO; companion has large pm\\
162417+064152 & 246.070533& $6.697785$      & 2  &  18.44   & 2.4              & C & similar color p-l; companion has large pm\\
163113-171407 & 247.804331& $-17.235305$    & 2  &  17.59   & 3.0              & C & similar color p-l; companion has large pm\\
163533+205229 & 248.886788& $20.87476$      & 2  &  18.49   & 2.1              & C & similar color p-l; companion has large pm\\
163614+094317 & 249.056552& $9.721352$      & 2  &  19.21    & 2.2              & C & similar color p-l; companion has large pm\\
163959-210652 & 249.995488& $-21.114331$    & 2  &  18.84   & 1.7              & C & similar color p-l; companion has large pm\\
164304+754120 & 250.765894& $75.688987$     & 2  &  18.08   & 3.5              & C & similar color p-l; companion has large pm\\
164552+152025 & 251.465796& $15.340377$     & 2  &  19.04   & 2.2               & B & similar color p-l; companion has large pm\\
170002+250336 & 255.009344& $25.060094$     & 3  &  18.50   & 4.7              & C & similar color p-l; 2 components have large pm\\
170024+005815 & 255.099997& $0.970862$      & 2  &  16.44   & 1.6              & C & similar color p-l; companion has large pm\\
170514+331637 & 256.307089& $33.276899$     & 2  &  18.98   & 2.2              & C & similar color p-l; includes SDSS z=2.224 QSO; companion has large pm\\
170516+251533 & 256.316261& $25.259123$     & 2  &  18.34   & 2.1              & C & similar color p-l; companion has large pm\\
170602+270515 & 256.509411& $27.087583$     & 2  &  17.31   & 2.3              & C & similar color; one component has large pm, the other one large AEN\\
170817+325311 & 257.072403& $32.886393$     & 2  &  18.38   & 2.1              & C & similar color p-l; companion has large AEN and pm\\
170858-030510 & 257.240718& $-3.086224$     & 2  &  17.60   & 2.5              & C & p-l; companion has large pm\\
170943+334304 & 257.427474& $33.717724$     & 2  &  19.13    & 3.3              & C & similar color p-l; both have large pm\\
171102+292951 & 257.757094& $29.497482$     & 2  &  17.92  & 2.2              & C & similar color p-l; includes SDSS z=1.329 QSO; SQLS candidate; \\
& &  &  &  & & & companion has large pm; no lensing object\\
172634+530300 & 261.639624& $53.050095$     & 2  &  18.76    & 1.3              & C & similar color p-l; includes white dwarf \citep{kleinman13}\\
173152+743615 & 262.968365& $74.604272$     & 3  &  16.60   & 5.9             & C & p-l (one has large p and pm) + red inner component\\
173316+084954 & 263.31709 & $8.831643$      & 2  &  17.78   & 2.4              & C & similar color p-l; companion has large pm\\
173509+094022 & 263.787393& $9.672832$      & 2  &  17.07  & 2.4              & C & similar color p-l; companion has large pm\\
173703+271724 & 264.262899& $27.290003$     & 2  &  18.37   & 2.5             & C & p-l; companion has large pm\\
173820+041756 & 264.581302& $4.298981$      & 2  &  18.88   & 2.0              & C & similar color p-l; companion has large pm\\
173905+120306 & 264.77013 & $12.051664$     & 2  &  17.77   & 2.8               & C & similar color p-l; companion has large pm\\
173915+112257 & 264.813269& $11.382484$     & 2  &  18.84    & 3.4              & C & similar color p-l; companion has large pm\\
174006+221101 & 265.024352& $22.183576$     & 2  &  17.45    & 1.8              & C & similar color p-l; includes z=1.406 QSO \citep{healey08}; \\
& &  &  &  & & & companion has large pm\\
174154+333616 & 265.474939& $33.604416$     & 3  &  16.54    & 8.5              & C & p-l; outer components have large p and pm\\
174213+402717 & 265.55245 & $40.454758$     & 2  &  18.30   & 3.1              & C & similar color p-l; companion has large pm\\
175243+093822 & 268.179389& $9.639313$      & 2  &  18.29   & 2.6              & C & similar color p-l; companion has large pm\\
175826+191732 & 269.608868& $19.292361$     & 2  &  17.57   & 2.5              & C & similar color p-l; companion has large pm\\
180257+244143 & 270.737205& $24.695406$     & 3  &  17.27    & 4.3              & C & similar color p-l (one companion has large pm) + red central component\\
180901+160103 & 272.254121& $16.017515$     & 2  &  18.59   & 3.1             & C & similar color p-l; companion has large pm\\
181045+742546 & 272.686785& $74.429515$     & 2  &  18.39    & 2.0              & C & similar color p-l; companion has large pm\\
181400+705410 & 273.499637& $70.902881$     & 2  &  17.69   & 2.4              & C & similar color p-l; companion has large pm\\
182159+275657 & 275.494183& $27.949111$     & 2  &  18.43   & 2.4              & C & similar color p-l; companion has large pm\\
182301+500140 & 275.753046& $50.027664$     & 2  &  17.91   & 2.0             & C & p-l; companion has large pm\\
183204+491637 & 278.015957& $49.276889$     & 2  &  18.06   & 1.9              & C & similar color p-l; companion has large pm\\
183852+520350 & 279.718445& $52.063814$     & 2  &  18.13    & 3.1              & C & p-l; companion has large pm\\
183916+454103 & 279.818168& $45.684238$     & 2  &  18.70   & 3.0              & C & similar color p-l; 4C 45.38, z=0.958 QSO; companion has large pm\\
184256+442102 & 280.733259& $44.350567$     & 2  &  18.33   & 3.1              & C & similar color p-l; companion has large pm\\
185008+441126 & 282.533367& $44.190435$     & 2  &  17.64   & 2.7               & C & similar color p-l; companion has large pm\\
185824+475553 & 284.600174& $47.931329$     & 3  &  18.54   & 3.6              & B & p-l (one component has large pm), red inner component\\
190003+522319 & 285.012245& $52.388677$     & 3  &  18.18   & 2.8              & C & similar color p-l (one component has large pm) + red inner component\\
190433+575031 & 286.139132& $57.841829$     & 2  &  19.00   & 3.2              & C & similar color p-l; companion has large pm\\
\hline
\end{tabular}
\\ 
\end{minipage}
\end{table*}
\normalsize

\begin{table*}
  \scriptsize
 \centering
 \begin{minipage}{\linewidth}
 \contcaption{}
  \begin{tabular}{@{}lrrclccl@{}}
  \hline 
Name [PS1~J...] & $\alpha$ & $\delta$ & \#Comp & $i$ & Sep. [\arcsec] & Rank & notes \\ 
 \hline
 192457+492126 & 291.239533& $49.357218$     & 2  &  18.71   & 2.8              & C & similar color p-l; companion has large pm\\
195629-064134 & 299.121219& $-6.692813$     & 2  &  19.12   & 1.6              & C & similar color p-l; companion has large pm\\
 200550-030100 & 301.456704& $-3.016733$     & 2  &  18.49   & 3.7              & C & similar color p-l; companion has large pm\\
 201810-022908 & 304.540147& $-2.485511$     & 2  &  18.25   & 1.8              & C & similar color p-l; companion has large pm\\
202339-290706 & 305.91091 & $-29.1182$      & 2  &  17.93   & 2.0               & C & similar color p-l; companion has large pm\\
203106-122005 & 307.776472& $-12.334677$    & 3  &  17.59   & 3.1             & C & similar color p-l (one component has large pm) + red companion\\
204541+122718 & 311.419538& $12.454995$     & 2  &  17.86   & 3.0             & C & similar color p-l; companion has large pm\\
204628-120049 & 311.615311& $-12.01355$     & 2  &  19.12   & 1.9             & C & similar color p-l; companion has large pm\\
210519+161334 & 316.330544& $16.226221$     & 2  &  18.71   & 1.7              & C & similar color p-l; companion has large pm\\
210820+122340 & 317.08394 & $12.394343$     & 2  &  17.89   & 1.8              & C & similar color p-l; one component has large p\\
211017+050707 & 317.571284& $5.118593$      & 2  &  18.60   & 2.8              & C & similar color p-l; companion has large pm\\
211945+153713 & 319.938477& $15.620234$     & 2  &  17.58   & 2.9              & C & similar color p-l; companion has large pm\\
212753+085302 & 321.972353& $8.883872$      & 2  &  18.50   & 2.2              & C & similar color p-l; companion has large pm\\
213147-030935 & 322.946983& $-3.159735$     & 3  &  18.52    & 6.9              & C & similar color p-l (one has large pm) + extended inner component\\
213707+124621 & 324.279402& $12.772593$     & 2  &  18.67   & 1.8              & C & similar color p-l; BL Lac \citep{d'abrusco14}; companion has large pm\\
214102+265252 & 325.257922& $26.881249$     & 2  &  17.78   & 2.8              & C & p-l; companion has large pm\\
214210+255233 & 325.543115& $25.875914$     & 2  &  16.95   & 2.4              & C & similar color p-l; companion has large pm\\
214248+290427 & 325.698926& $29.074187$     & 2  &  17.38   & 3.6              & C & similar color p-l; X-ray source \citep{d'abrusco14}; \\
& &  &  &  & & & companion has large p and pm\\
214605+264507 & 326.52051 & $26.75202$      & 2  &  19.2      & 2.3               & C & similar color p-l; companion has large pm\\
215502+190303 & 328.756839& $19.050739$     & 2  &  16.86   & 2.1              & B & similar color p-l; QSO+star (NTT run 0100.A-0297(A), PI. T. Anguita)\\
220822-142722 & 332.093734& $-14.455987$    & 2  &  17.25     & 2.1              & C & similar color p-l; companion has large p and pm\\
222238+354225 & 335.656449& $35.707081$     & 2  &  18.30   & 2.1              & C & similar color p-l; companion has large pm\\
222611-282413 & 336.547769& $-28.403508$    & 2  &  19.18   & 2.7              & C & similar color p-l; includes z=0.016 galaxy \citep{maddox90}; \\
& &  &  &  & & & companion has large pm\\
223604+221604 & 339.015242& $22.267863$     & 2  &  17.98   & 3.1              & C & similar color p-l; companion has large pm\\
223713+245120 & 339.304497& $24.85563$      & 2  &  18.26   & 1.7              & B & similar brightness p-l; companion has large pm\\
223831+140027 & 339.629583& $14.007554$     & 2  &  19.50   & 3.1              & B & similar color p-l + extended? HS 2236+1344 (blue compact galaxy); \\
& &  &  &  & & & both have large AEN, no other Gaia data\\
230258-281314 & 345.740301& $-28.220566$    & 2  &  18.24    & 1.7              & C & similar color p-l; QSO+star \citep{lemon18}\\
231209+203543 & 348.036702& $20.595139$     & 2  &  19.16   & 3.1              & C & p-l + extended; companion has large pm\\
231313+194722 & 348.302961& $19.7895$       & 2  &  17.63   & 3.2              & C & similar color p-l; companion has large pm\\
231445+303530 & 348.687176& $30.591695$     & 2  &  17.20   & 3.0              & C & p-l; companion has large p and pm\\
232837+435308 & 352.152836& $43.885431$     & 2  &  17.33    & 2.1              & B & p-l; companion has large pm\\
233611-093523 & 354.043989& $-9.589647$     & 2  &  15.63   & 4.1              & C & similar color p-l; companion has large p and pm\\
233700+180520 & 354.249022& $18.088753$     & 2  &  17.63   & 3.7              & C & p-l; companion has large pm\\
234155+132902 & 355.480568& $13.483904$     & 2  &  18.82   & 3.2              & C & similar color p-l; includes SDSS z=0.729 QSO; companion has large pm\\
235351-053956 & 358.462667& $-5.665505$     & 3  &  16.5    & 6.2               & B & similar color p-l (companion has large p and pm) + red inner component; \\
& &  &  &  & & & QSO+star \citep{williams17}\\
\hline
\end{tabular}
\\ 
{\footnotesize The systems above the horizontal line are confirmed lenses or quasar pairs. The ones below are candidates ruled out either due to their $Gaia$-based properties, or due to spectroscopic results from the literature. The table structure is the same as in Table \ref{tab:cand}. ``NIQ'' stands for nearly identical quasars.}
\label{tab:candappend}
\end{minipage}
\end{table*}
\normalsize



\bsp	
\label{lastpage}
\end{document}